\documentstyle[]{mn}
\newif\ifAMStwofonts

\ifoldfss
  \newcommand{\rmn}[1] {{\rm #1}}

  \ifCUPmtlplainloaded \else
    \NewTextAlphabet{textbfit} {cmbxti10} {}
    \NewTextAlphabet{textbfss} {cmssbx10} {}
    \NewMathAlphabet{mathbfit} {cmbxti10} {} 
    \NewMathAlphabet{mathbfss} {cmssbx10} {} 
  \fi
  \ifAMStwofonts
    \ifCUPmtlplainloaded \else
      \NewSymbolFont{upmath} {eurm10}
      \NewSymbolFont{AMSa} {msam10}
      \NewMathSymbol{\upi}     {0}{upmath}{19}
      \NewMathSymbol{\umu}     {0}{upmath}{16}
      \NewMathSymbol{\upartial}{0}{upmath}{40}
      \NewMathSymbol{\leqslant}{3}{AMSa}{36}
      \NewMathSymbol{\geqslant}{3}{AMSa}{3E}

      \let\leq=\leqslant 
       
    \fi
  \fi
\fi 

\ifnfssone
  \newmathalphabet{\mathit}
  \addtoversion{normal}{\mathit}{cmr}{m}{it}
  \addtoversion{bold}{\mathit}{cmr}{bx}{it}
  \newcommand{\rmn}[1] {\mathrm{#1}}

  \newmathalphabet{\mathbfit} 
  \addtoversion{normal}{\mathbfit}{cmr}{bx}{it}
  \addtoversion{bold}{\mathbfit}{cmr}{bx}{it}
  \newmathalphabet{\mathbfss} 
  \addtoversion{normal}{\mathbfss}{cmss}{bx}{n}
  \addtoversion{bold}{\mathbfss}{cmss}{bx}{n}
  \ifAMStwofonts
    \ifCUPmtlplainloaded \else
      \UseAMStwoboldmath
      \makeatletter
      \new@mathgroup\upmath@group
      \define@mathgroup\mv@normal\upmath@group{eur}{m}{n}
      \define@mathgroup\mv@bold\upmath@group{eur}{b}{n}
      \edef\UPM{\hexnumber\upmath@group}
      \new@mathgroup\amsa@group
      \define@mathgroup\mv@normal\amsa@group{msa}{m}{n}
      \define@mathgroup\mv@bold\amsa@group{msa}{m}{n}
      \edef\AMSa{\hexnumber\amsa@group}
      \makeatother
      \mathchardef\upi="0\UPM19
      \mathchardef\umu="0\UPM16
      \mathchardef\upartial="0\UPM40
      \mathchardef\leqslant="3\AMSa36
      \mathchardef\geqslant="3\AMSa3E

      \let\leq=\leqslant 

    \fi
  \fi
\fi 

\ifnfsstwo
  \newcommand{\rmn}[1] {\mathrm{#1}}

  \DeclareMathAlphabet{\mathbfit}{OT1}{cmr}{bx}{it}
  \SetMathAlphabet\mathbfit{bold}{OT1}{cmr}{bx}{it}
  \DeclareMathAlphabet{\mathbfss}{OT1}{cmss}{bx}{n}
  \SetMathAlphabet\mathbfss{bold}{OT1}{cmss}{bx}{n}
  \ifAMStwofonts
    \ifCUPmtlplainloaded \else
      \DeclareSymbolFont{UPM}{U}{eur}{m}{n}
      \SetSymbolFont{UPM}{bold}{U}{eur}{b}{n}
      \DeclareSymbolFont{AMSa}{U}{msa}{m}{n}
      \DeclareMathSymbol{\upi}{0}{UPM}{"19}
      \DeclareMathSymbol{\umu}{0}{UPM}{"16}
      \DeclareMathSymbol{\upartial}{0}{UPM}{"40}
      \DeclareMathSymbol{\leqslant}{3}{AMSa}{"36}
      \DeclareMathSymbol{\geqslant}{3}{AMSa}{"3E}

      \let\leq=\leqslant 

    \fi
  \fi
\fi 

\ifCUPmtlplainloaded \else
  \ifAMStwofonts \else 
    \def\upi{\pi}
    \def\umu{\mu}
    \def\upartial{\partial}
  \fi
\fi

\newcommand{\unit}[1]
        {{\mbox{\rm\,\,#1}}}

\newcommand{\m}
        {\unit{m}}

\newcommand{\persqdeg}
        {\unit{deg$^{-2}$}}
\newcommand{\sqdeg}
        {\unit{deg$^2$}}

\newcommand{\grs}
        {GRS}
\newcommand{\grss}
        {GRSs}

\newcommand{\IRAS}
        {{\em IRAS}}
\newcommand{\CNOC}
        {CNOC}
\newcommand{\TdF}
        {2dF}

\newcommand{\LCRS}
        {LCRS}
\newcommand{\MRP}
        {MRSP}

\newcommand{\SDSS}
        {SDSS}

\newcommand{\CfA}
        {CfA}

\newcommand{\ESP}
        {ESP}
\newcommand{\APM}
        {APM}
\newcommand{\psf}
        {PSF}

\newcommand{\samethanks}
	{{\Huge $^\star$}}

\newcommand{\twoint}
        {\int \!\!\! \int}

\newcommand{\vect}[1]
        {\mbox{\boldmath ${#1}$}}

\newcommand{\etc}
	{etc.}
\newcommand{\etal}
	{et al.}
\newcommand{\eg}
	{e.g.}
\newcommand{\cf}
	{c.f.}
\newcommand{\ie}
	{i.e.}
\newcommand{\eq}[1]
	{equation~(\ref{equation:#1})}
\newcommand{\eqs}[1]
	{equations~(\ref{equation:#1})}

\newcommand{\sect}[1]
	{Section~\ref{section:#1}}

\newcommand{\sects}[1]
        {Sections~\ref{section:#1}}
\newcommand{\Sect}[1]
        {Section~\ref{section:#1}}

\newcommand{\tabl}[1]
        {{\mbox Table~\ref{table:#1}}}

\newcommand{\fig}[1]
	{Fig.~\ref{figure:#1}}

\newcommand{\Fig}[1]
        {Fig.~\ref{figure:#1}}

\newlength{\singlefigureheight}
\newlength{\doublefigureheight}
\newlength{\triplefigureheight}
\newlength{\squarefigureheight}
\newcommand{\AaA}
        {A\&A}

\newcommand{\AJ}
        {AJ}
\newcommand{\ApJ}
        {ApJ}
\newcommand{\ApJS}
        {ApJS}

\newcommand{\ARAA}
        {ARA\&A}

\newcommand{\MNRAS}
        {MNRAS}
\newcommand{\PhilTransA}
        {Phil.\ Trans.\ of the Royal Soc.\ A}

\newcommand{\PASP}
        {PASP}

\newcommand{\Science}
        {Science}

\newcommand{\omo}
        {\Omega_{\rmn m_0}}

\newcommand{\olo}
        {\Omega_{\Lambda_0}}

\newcommand{\FL}
        {FL}

\begin{document}

\title[Lensing in redshift surveys]
{Using galaxy redshift surveys to detect gravitationally-lensed quasars}

\author[D.\ J.\ Mortlock and R.\ L.\ Webster]
       {
        Daniel J.\ Mortlock$^{1,2,3}$\thanks{
		E-mail: mortlock@ast.cam.ac.uk (DJM);
		rwebster@physics. unimelb.edu.au (RLW)}
	and Rachel L.\ Webster$^1$\samethanks\ \\
        $^1$School of Physics, The University of Melbourne, Parkville,
        Victoria 3052, Australia \\
        $^2$Astrophysics Group, Cavendish Laboratory, Madingley Road,
        Cambridge CB3 0HE, U.K. \\
        $^3$Institute of Astronomy, Madingley Road, Cambridge
        CB3 0HA, U.K. \\
       }

\date{
Accepted. 
Received; in original form 2000 April 19}

\pagerange{\pageref{firstpage}--\pageref{lastpage}}
\pubyear{2000}

\label{firstpage}

\maketitle

\begin{abstract}
Gravitationally-lensed quasars 
can be discovered as a by-product of galaxy redshift surveys.
Lenses discovered spectroscopically in this way should require
less observational effort per event 
than those found in dedicated lens surveys.
Further, the lens galaxies should be relatively nearby,
facilitating a number of detailed observations that 
are impossible for the more common high-redshift lenses. 
This is epitomised by the wide range of results that have 
been obtained from Q~2237+0305, which was 
discovered as part of the Center for Astrophysics redshift survey,
and remains the only quasar lens discovered in this way.
The likelihood of this survey yielding a lens is calculated to be  
$\sim 0.03$, which is an order of magnitude larger than
previous estimates due to two effects.
Firstly, the quasar images themselves increase the observed
flux of the lens, so that lens galaxies up to a magnitude fainter
than the nominal survey limit must be included in the calculation.
Secondly, it is possible for lensed quasars
with extremely faint deflectors to
enter the survey due to the extended morphology of the multiple images.
Extrapolating these results to future surveys, the 
2 degree Field
galaxy redshift survey should contain between 10 and 50 lenses
and the
Sloan Digital Sky Survey should yield
between 50 and 300 lenses, depending
on the cosmological model and the observing conditions.
\end{abstract}


\section{Introduction}
\label{section:intro_grs}

Gravitationally-lensed quasars are extremely valuable for 
a number of reasons.
The frequency of lensing events is a sensitive probe 
of the cosmological constant (\eg\ Turner 1990; Kochanek 1995, 1996a)
and the density of the universe (Kochanek 1995; Mortlock \& Webster 2000a).
Individual lenses can be used to constrain the mass
distribution of the deflector 
(\eg\ Chen, Kochanek \& Hewitt 1995; 
Chae, Turnshek \& Khersonsky 1998; Keeton, Kochanek \& Falco 1998),
the mass of galactic halo objects 
(\eg\ Schmidt \& Wambsganss 1998; Wyithe, Webster \& Turner 2000b),
extinction in the lens galaxy 
(\eg\ Malhotra, Rhoads \& Turner 1997; Falco \etal\ 1999),
and Hubble's constant 
(\eg\ Refsdal 1964; Grogin \& Narayan 1996; Kundi\'{c} \etal\ 1997).
Further, gravitational lensing is one of the only means by which
these properties of high-redshift galaxies can be probed.

Conversely, lenses with low-redshift deflectors are also
particularly useful, but for rather different reasons.
The best illustration of this is
Q~2237+0305 (Huchra \etal\ 1985),
a quadruply-imaged quasar at a redshift of 1.69, seen
through the bulge of a spiral galaxy at a redshift of 0.04.
The proximity of the lens galaxy
is such that it is detectable in H\,{\sc i} (Barnes \etal\ 1999), has
an optically measured velocity dispersion
(Foltz \etal\ 1992), and can be resolved on scales of $\la 100$ pc by
the {\em Hubble Space Telescope}. These results are
unremarkable in themselves, but can be combined with lensing
constraints to provide a wealth of information about both the
galaxy and the lensed quasar.
The macroscopic properties of the images allow accurate
determinations of the mass distribution and mass-to-light
ratio of the galaxy (\eg\ Yee 1988; Kent \& Falco 1988)
and, combined with their observed position angles,
the disk and bar can be weighed independently
(Schmidt, Webster \& Lewis 1998).
Microlensing of the quasar images by compact objects within
the galaxy permits the measurement of their mean mass
(Lewis \& Irwin 1996; Wyithe, Webster \& Turner 2000a),
as well as a determination of
the continuum source size of the quasar 
(Wambsganss, Paczy\'{n}ski \& Schneider 1990;
Rauch \& Blandford 1992;
Wyithe \etal\ 2000c), and even the
transverse motion of the galaxy relative to the line-of-sight
to the quasar (Wyithe, Webster \& Turner 1999).

After Q~2237+0305, 
the second nearest deflector is the redshift 0.11 elliptical galaxy
that lenses the radio source MG~1549+3047 (Leh\'{a}r \etal\ 1993).
Both the mass
distribution and velocity dispersion of the galaxy have been
measured (Leh\'{a}r \etal\ 1996), but the greater distance 
to the lens and 
less optimal source alignment place limits on the inferences
that can be made from this system.

It is clearly desirable to have a larger sample of lenses with
low-redshift deflectors.
Each would have the potential to provide as much insight
as Q~2237+0305 and a sufficiently large sample could be used to 
extrapolate from the above results to the galaxy and source
populations (\eg\ Keeton \etal\ 1998).
The limitations of the various lens survey techniques are 
discussed in \sect{surveys}, with particular emphasis on the 
use of galaxy redshift survey (\grs) spectra as the primary data.
\Sect{num counts} describes a simple model of the various 
populations of objects that determine the viability of 
this method.
In \sect{results} specific predictions are made for 
various existing and planned redshift surveys,
and the results summarised in \sect{conc}.

\section{Lens surveys}
\label{section:surveys}

Lensed quasars are very rare,
with less than 50 known to date (\eg\ Keeton \& Kochanek 1996).
At most 1 in $\sim 100$ (and probably closer to 1 in $\sim$1000)
quasars brighter 
than $m \simeq 19$ are lensed, which implies a surface 
density of $\la 0.1\persqdeg$.
This is several orders of magnitude lower than the
integrated number counts of stars 
($\sim$1000$\persqdeg$; \eg\ Jones \etal\ 1991),
galaxies ($\sim 100\persqdeg$; \eg\ Maddox \etal\ 1990b), 
and unlensed quasars 
($\sim 10\persqdeg$; \eg\ Hewett, Foltz \& Chaffee 1995)
to the same limit.
Thus lensed quasars represent between 0.001 per cent and
0.01 per cent of `bright' optical sources.
Quasars (and hence lenses) represent a larger fraction of the radio source 
counts, and a number of large radio lens surveys exist
(\eg\ Patnaik \etal\ 1992; Myers \etal\ 1995).
Unfortunately the lack of knowledge about the luminosity
function and redshift distribution of the sources limits the 
power of any statistical results (Kochanek 1996b).

Compounding these problems, considerable observational
effort is required to both identify candidate lenses 
and to 
then confirm their identity.
Most are discovered in lens surveys
(\eg\ Burke, Leh\'{a}r \& Connor 1992; 
Crampton, McClure \& Fletcher 1992; 
Maoz \etal\ 1992; Patnaik \etal\ 1992;
Maoz \etal\ 1993a,b; Surdej \etal\ 1993; 
Yee, Filippenko \& Tang 1993;
Falco 1994; Kochanek, Falco \& Schild 1995;
Jackson \etal\ 1995;
Jaunsen \etal\ 1995) which are based
on existing quasar (or extra-Galactic radio-source) samples.
These catalogues in themselves represent a great deal of work,
as initial multi-colour selection of candidates must then be
followed by confirmation spectroscopy of each potential quasar.
However there are now a number of large quasar
samples in existence (\eg\ Boyle, Shanks \& Peterson 1988; 
Warren, Hewett \& Osmer 1991;
Hewitt \& Burbidge 1993; Hewett \etal\ 1995)
or in preparation (\eg\ Loveday \& Pier 1998; Boyle \etal\ 1999a,b), so 
any new lens survey can proceed from these object lists.
Under this assumption, a lens survey then requires high-resolution
imaging of between 100 and 1000 fields (\ie\ one for each quasar) 
per lens discovery. (Further observations of any probable lenses
are usually required to establish the lensing interpretation
beyond doubt, but these represent only a small part of the overall
task.)

The efficiency of a lens survey can be increased by the use of 
morphological selection criteria, for instance choosing only
highly-elliptical sources for further investigation.
Kochanek (1991) found that, whilst the efficiency of such 
searches is quite high, 
they can be very incomplete, finding only $\sim 20$ per cent of
all lenses. This conclusion was consistent with the
results of
the Automatic Plate Measuring (\APM) lens survey 
(Webster, Hewett \& Irwin 1988),
which contains no lenses amongst $\sim 2500$ elliptical sources
with quasar-like colours.

Lens surveys also tend to miss some of the most interesting
multiply-imaged quasars.
These include those
with high extinction and reddening from dust along the line-of-sight,
and those with low-redshift lens galaxies.
As discussed above, the latter can be very valuable indeed,
but cannot be found in lens surveys they
are not likely to enter the quasar surveys in the first place,
due to both their morphologies (appearing simply to be nearby galaxies
at all but the highest resolutions) and colours (which are likely to
be galaxy-like).
Lenses with low-redshift deflectors
can only be found by searching for quasar emission
lines in the spectra obtained in \grss.

Indeed, Q~2237+0305 
was discovered as a lens when it was noticed that one of the galaxies in the 
Center for Astrophysics (\CfA) redshift survey (Geller \& Huchra 1989) had 
an unusual spectrum.
Follow up spectroscopy and imagining confirmed that the object
was in fact a lens.
Whilst Q~2237+0305 represents,
a posteriori, an unlikely alignment\footnote{Kochanek (1992) estimated
the probability of a lens being discovered in the \CfA\ survey was 
$\sim 0.003$, but see \sect{num_grs}.}, it is 
quite probable that other lenses will be discovered in redshift
surveys, given that at least 1 in 10$^6$ 
redshifts will yield a lens (Kochanek 1992).
Aside from the \CfA\ survey 
(which includes $\sim 1.5 \times 10^4$ objects),
several other large 
\grss\ have already been completed.
The Las Campanas Redshift Survey (\LCRS; Shectman \etal\ 1996) and
the {\em InfraRed Astronomical Satellite}
(\IRAS) 1.2 Jy
survey (Fisher \etal\ 1995)
have both measured upwards of $10^4$ galactic redshifts,
and so may contain a few lenses between them.
With $\sim 9 \times 10^5$ galaxies,
the Muenster Redshift Project
(\MRP; Schuecker, Ott \& Seitter 1996)
dwarfs the other completed surveys, 
but is probably less useful as a source of lenses,
the redshifts having been measured from 
low-dispersion objective prism spectra,
as opposed to higher-resolution spectrographic data.
In the future,
the 2 degree Field (\TdF) \grs\ (\eg\ Colless 1999; Folkes \etal\ 1999)
will include  
$\sim 2.5 \times 10^5$ redshift measurements, and the 
Sloan Digital Sky Survey (\SDSS; \eg\ Szalay 1998; Loveday \& Pier 1998)
will contain about one million galaxies. 
The \SDSS\ could yield
`several tens' of spectroscopically-discovered lenses
(Kochanek 1992),
although it will also include
high-resolution imaging, and thus should find a large 
number of lenses `directly'.
Importantly, these estimates are really only lower bounds on the number
of lenses in \grss, as previous calculations did not take into account
two important observational effects.

The first is similar to the standard lensing 
magnification bias (\eg\ Turner 1980), in which the number of lenses in a 
magnitude-limited sample is increased as 
intrinsically faint lensed sources
are preferentially magnified into the survey.
In a galaxy sample there is the additional effect that
the magnified quasar images tend to be merged with the lens
galaxy, making the composite object appear brighter. In this way 
(potential) lens galaxies up to one mag fainter than the survey limit
can enter a \grs\ sample.
As shown in \sect{num counts}, this effect can increase 
the number of lenses in a redshift survey by up to an
order of magnitude.

The second effect is complementary to this,
as it only becomes important if the lens galaxy is much
fainter than the quasar images (which is true of most 
lenses discovered to date).
Such a lens would appear as several point sources,
but might
enter a \grs\ by virtue of appearing to be a single extended
source in the low-resolution data from which survey
catalogues are generated. 
This is a potentially important possibility, but it
is highly-dependent on the details of 
the survey selection effects and point-spread function (\psf),
as discussed in \sects{num counts} and \ref{section:results}.

Assuming that a number of the objects in a \grs\ are 
lenses, and are hence observed spectroscopically as a matter of
course, a minimal number of confirmation observations are required to 
find them. This method is only practical if it is possible
to differentiate between quasar spectra and galaxy spectra (as well
as those of stars, white dwarfs, \etc), which, fortunately, is
reasonably straight-forward, due to their prominent, broad
emission lines. 
Other than the lenses,
the only objects that will be selected in this way
are any unlensed quasars that 
appear extended in the parent survey.
These must also be re-imaged, and so the number of 
unlensed quasars in a \grs\ determines 
the efficiency of this type of lens search.

\section{Source populations}
\label{section:num counts}

\begin{figure}
\includegraphics{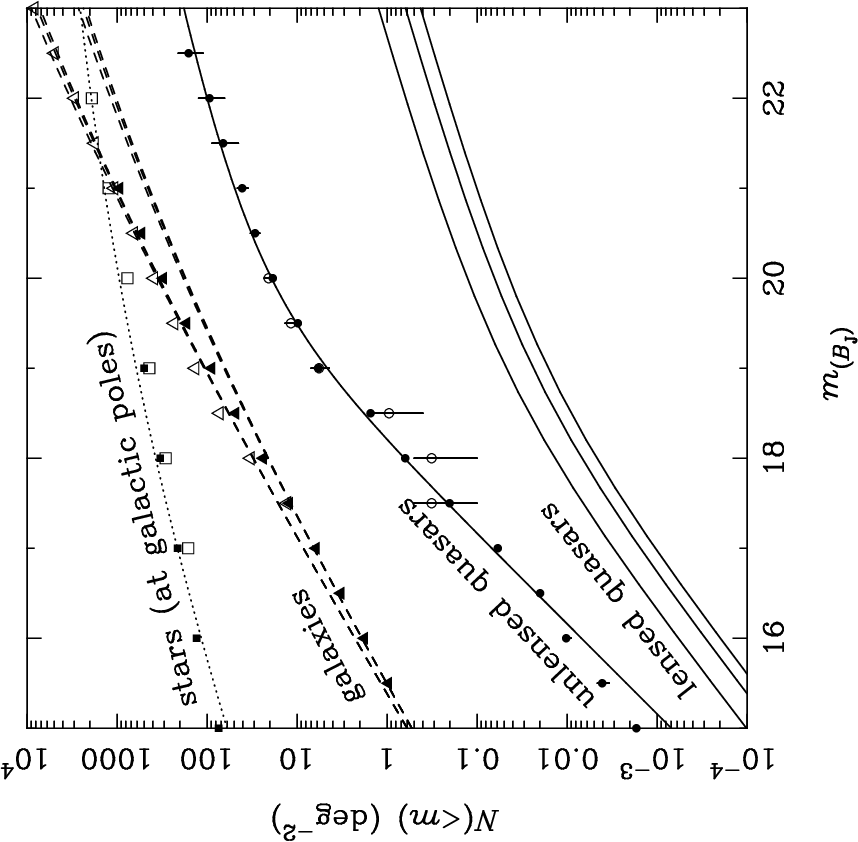}
\vspace{\squarefigureheight}
\caption{Cumulative number counts of stars (at the Galactic poles;
dotted line), galaxies
(dashed lines), unlensed quasars (upper solid line)
and lensed quasars (lower solid lines) as a function of
$B_{\rmn J}$ magnitude.
The observed stellar number counts are taken from
Stobie \& Ishida (1987; filled squares) and 
Jones \etal\ (1991; empty squares);
the fit (from Bahcall \& Soneira 1990) is described in \sect{stars}.
The observed galaxy counts are from 
the \APM\ galaxy survey
(Maddox \etal\ 1990b; filled triangles)
and
Jones \etal\ (1991; open triangles).
The no-evolution galaxy models (\sect{galaxies}) are given for
$\Omega_{\rm m_0} = 1.0$ and $\Omega_{\Lambda_0} = 0.0$;
$\Omega_{\rm m_0} = 0.3$ and $\Omega_{\Lambda_0} = 0.0$;
and $\Omega_{\rm m_0} = 0.3$ and $\Omega_{\Lambda_0} = 0.7$,
although the number counts
are almost independent of the cosmological model.
The $K$-corrections
of Kinney \etal\ (1996) are used for the lower triplet of dashed lines,
whereas passive luminosity evolution is assumed to cancel the 
$K$-corrections for the upper triplet of dashed lines.
The observed number counts of unlensed quasars
are taken from the compilation of data
(the sum of the $z \leq 2.2$ and $z > 2.2$ samples)
presented by Hartwick \& Schade (1990; filled circles),
and the Fornax spectroscopic survey (Drinkwater \etal\ 1999; open 
circles);
the fit given in \sect{quasars}.
All lenses (\ie\ irrespective of image separation or flux ratio)
are included in the three lower solid lines,
which assume:
$\Omega_{\rm m_0} = 1.0$ and $\Omega_{\Lambda_0} = 0.0$ (lower line);
$\Omega_{\rm m_0} = 0.3$ and $\Omega_{\Lambda_0} = 0.0$ (middle line);
and
$\Omega_{\rm m_0} = 0.3$ and $\Omega_{\Lambda_0} = 0.7$ (upper line).}
\label{figure:num counts}
\end{figure}

The difficulty of finding gravitational lenses is principally
a function of their rarity, so any assessment of search methods
must include an estimate of the total number of 
sources which must be investigated per lens detection.
The source populations considered are:
stars (\sect{stars});
galaxies (\sect{galaxies});
unlensed quasars (\sect{quasars});
and lensed quasars (\sect{lensed quasars}).
Each population is defined by
$N(<m)$, the 
number of objects per unit solid angle brighter than magnitude\footnote{The
analysis presented here is reasonably generic, but in general 
$B_{\rmn J}$ or $B$ magnitudes are considered.}
$m$, which is 
shown for each of the four classes of objects in 
\fig{num counts}.

In this simple model, 
the morphology of an object is considered to be completely
defined by its ellipticity, 
\begin{equation}
\label{equation:e}
e = 1 - \left[
\frac{M_{xx} + M_{yy} - \sqrt{(M_{xx} - M_{yy})^2 + 4 M_{xy}}}
{M_{xx} + M_{yy} + \sqrt{(M_{xx} - M_{yy})^2 + 4 M_{xy}}}
\right]^{1/2},
\end{equation}
where the moments, $M$, are given by
\begin{equation}
M_{xx} = \langle (\theta_x - \langle \theta_x \rangle)^2 \rangle,
\end{equation}
\begin{equation}
M_{yy} = \langle (\theta_y - \langle \theta_y \rangle)^2 \rangle
\end{equation}
and 
\begin{equation}
M_{xy} = \langle (\theta_y - \langle \theta_y \rangle)
(\theta_x - \langle \theta_x \rangle) \rangle.
\end{equation}
The angled brackets denote expectation values,
defined by 
\begin{equation}
\label{equation:expectation}
\langle h \rangle = \twoint
h(\vect{\theta}) f(\vect{\theta}) \, {\rmn d}\theta^2,
\end{equation}
where the integrals extend over the whole sky, 
and $f(\vect{\theta})$ is the observed surface brightness
of the object.
The shape distribution of each population is then defined by
${\rmn d}p/{\rmn d}e|_{<m}$, the ellipticity distribution
of all objects brighter than $m$.
This is a significant simplification, the limitations
of which are discussed where relevant. More complex methods 
(\eg\ Jarvis \& Tyson 1981; Godwin, Metcalfe \& Peach 1983; 
Maddox \etal\ 1990a)
provide more powerful methods of separating galaxies from stellar 
images, but are also more closely tied to individual surveys.
The analysis presented here is reasonably 
general; Mortlock \& Webster (2000c) present a more detailed 
and realistic simulation lensing in the \TdF\ \grs,
which uses the sophisticated \APM\ star-galaxy separation algorithm.

\subsection{Stars}
\label{section:stars}

By far the most common bright point-like sources on the sky are
stars, and their sheer numbers
place limits on the efficiency of any search for lenses 
which includes point sources. Their density is 
not uniform across the sky, and thus it is preferable to 
perform any extra-Galactic survey away from the disk of the
Milky Way, most commonly towards the Galactic
poles. 

\subsubsection{Number counts}

The integrated number counts of stars in the $B$-band towards
the Galactic poles can be approximated (to within 
$\sim 15$ per cent) by 
\begin{equation}
\label{equation:n_stars}
N_{\rmn s} (< m) =
\end{equation}
\[
\mbox{} \frac{N_1 10^{0.027(m - m_{\rmn s1})}}
{\left[1 + 10^{- 0.124(m - m_{\rmn s1})}\right]^{3.1}}
+ \frac{N_2 10^{0.083(m - m_{\rmn s2})}}
{\left[1 + 10^{- 0.167(m - m_{\rmn s2})}\right]^{2.5}},
\]
where $N_1 \simeq 3.9 \times 10^7 \simeq 950 \persqdeg$, 
$m_{\rmn s1} \simeq 16.6$,
$N_2 \simeq 3.7 \times 10^7 \simeq 910 \persqdeg$, and 
$m_{\rmn s2} \simeq 18.0$ 
(Bahcall \& Soneira 1980).
This is shown as the dotted line in \fig{num counts}, 
along with the observed number counts (near the south
Galactic pole) from Stobie \& Ishida (1987) 
and Jones \etal\ (1991).

\subsubsection{Ellipticity distribution}

\begin{figure}
\includegraphics{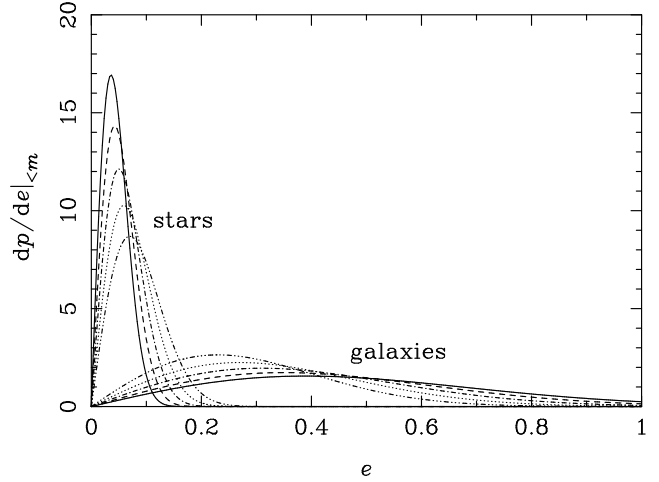}
\vspace{\singlefigureheight}
\caption{The ellipticity distributions of
stars (and other point-sources) and galaxies, as labelled.
The five distributions are for objects with
$m < 17$ (solid lines),
$m < 18$ (dashed lines),
$m < 19$ (dot-dashed lines),
$m < 20$ (dotted lines),
and
$m < 21$ (dot-dot-dot-dashed lines).
The distribution of stellar
ellipticities broadens and the
galactic ellipticities becomes narrower towards fainter magnitudes,
as the effects of noise begin to dominate.}
\label{figure:e_theory}
\end{figure}

Despite the fact that all but the closest stars are un-resolved,
most appear to have 
finite ellipticity due to noise effects. 
The \APM\ survey is considered as a representative 
example of the low-resolution data used to generate galaxy
survey catalogues, and so the stellar ellipticity distribution
was chosen to match this, as described in Mortlock (1999).
The functional form adopted was a two-dimensional Gaussian, given by
\begin{equation}
\label{equation:dpde_stars}
\left. \frac{{\rmn d}p_{\rmn s}}{{\rmn d} e} \right|_{< m}
= \frac{\pi e}{2 \langle e_{\rmn s} \rangle^2 (<m)} 
\exp\left[- \frac{\pi}{4} \left(\frac{e}{\langle e_{\rmn s} \rangle(< m)} 
\right)^2\right],
\end{equation}
with the mean ellipticity over the range of interest 
($m \la 22$)
well-approximated by
\begin{equation}
\langle e_{\rmn s} \rangle (< m)
= \frac{1}{20} \exp\left(\frac{m - 19}{6}\right).
\end{equation}
This distribution is shown in
\fig{e_theory}, demonstrating
both the variation with magnitude and the degree of overlap 
with the galactic ellipticity distribution
(as given in \sect{galaxies}).

\subsection{Galaxies}
\label{section:galaxies}

The galaxy population not only dominates the number counts 
of extended sources, but also determines the number of lensed
quasars, as the majority of deflectors are isolated galaxies.
Three galaxy types 
are considered: spirals
(S) and two classes of ellipticals (E and S0), with their relative 
numbers given by $\nu_{\rmn S} \simeq 0.69$;
 $\nu_{\rmn{E}} \simeq 0.12$;
and $\nu_{\rmn{S}0} \simeq 0.19$ (Postman \& Geller 1984).
The co-moving galaxy luminosity function of each type
is assumed to 
follow the same Schechter (1976) law, given by
\begin{eqnarray}
\label{equation:schechter_grs}
\frac{{\rmn d}n_{\rmn g}}{{\rmn d}M} & = &
- \frac{2}{5} \ln(10) \nu n_{\rmn g*} \nonumber \\
& \times &
10^{-2/5\, (M - M_*) (\alpha + 1)} 
\exp\left[- 10^{-2/5 \, (M - M_*)} \right],
\end{eqnarray}
where 
$n_{\rmn g*} = (0.016 \pm 0.003) h^3$ Mpc$^{-3}$ 
gives the number density of all galaxies,
$M_* = - 19.4 \pm 0.1$ in the $B$-band
and 
$\alpha = - 1.07 \pm 0.05$ (Efstathiou, Ellis \& Peterson \ 1988).

\subsubsection{Number counts}

Assuming \eq{schechter_grs} holds at all redshifts (\ie\ no evolution),
the differential galaxy number counts for each population are given by
\begin{equation}
\label{equation:dndmdz_gal}
\frac{{\rmn d} N_{\rmn g}}{{\rmn d}m}
= \frac{1}{4 \pi}
\int_0^\infty \frac{{\rmn d}V_0}{{\rmn d}z} 
\left. \frac{{\rmn d}n_{\rmn g}}{{\rmn d}M} \right|_{M = M(m, z)}
\, {\rmn d}z,
\end{equation}
where 
${\rmn d}V_0/{\rmn d}z$ is the
co-moving volume of a shell at redshift $z$
(\eg\ Carroll, Press \& Turner 1992).
Both the volumes and the distances are dependent on
the current normalised density of the universe, 
$\Omega_{\rm m_0}$,
and 
the normalised cosmological constant,
$\Omega_{\Lambda_0}$. 
It is this
dependence that allows both gravitational lensing statistics and number
counts to be useful as cosmological probes, although
the galactic number counts in the magnitude range considered
here are almost independent of the cosmological model.
The relationship between apparent and absolute magnitudes
is 
\begin{equation}
\label{equation:m_abs_z}
M (m, z) = m - 25 - 
5 \log \left[\frac{d_{\rmn L}(0, z)}{{\rmn 1 \, Mpc}}\right]
- K(z) - E(z),
\end{equation}
where $d_{\rmn L}(0, z)$ is the luminosity distance
(\eg\ Carroll \etal\ 1992),
$K(z)$ the $K$-correction 
and $E(z)$ the evolutionary correction, 
both of which are type-dependent. 
Two models are considered here: 
a non-evolving model with $E(z) = 0$ and 
$K$-corrections as measured by Kinney \etal\ (1996);
and 
one in which the 
$K$-correction is cancelled out by the effects of passive 
luminosity evolution [\ie\ $E(z) = - K(z)$].
Out to moderate redshifts the latter model gives a much better 
fit to the galaxy number counts (\eg\ \fig{num counts};
Maddox \etal\ 1990b), and so is adopted as the default model.

The cumulative number counts for each population 
are obtained by integrating
\eq{dndmdz_gal} to give
\begin{equation}
\label{equation:n_g_m}
N_{\rmn g} (< m) = \int_{- \infty}^{m} \int_0^\infty
\frac{{\rmn d}V_0}{{\rmn d}z}
\left. \frac{{\rmn d}n_{\rmn g}}{{\rmn d}M} \right|_{M = M(m^\prime, z)}
\,{\rmn d}z \, {\rmn d}m^\prime ,
\end{equation}
which is shown as the dashed lines in \fig{num counts}.

\subsubsection{Ellipticity distribution}

Galaxies are the archetypal extended astronomical object, but the
ellipticity distribution of point sources (\sect{stars})
is sufficiently broad that there is some overlap.
This is especially true for fainter objects -- 
the ellipticity distributions of both stars and galaxies
approach a noise-dominated distribution of 
intermediate mean ellipticity.
A two-dimensional Gaussian is used to parameterise
the distribution of galactic ellipticities 
[\cf\ \eq{dpde_stars}], so that
\begin{equation}
\label{equation:dp_gde}
\frac{{\rmn d}p_{\rmn g}}{{\rmn d}e} = \frac{\pi e}{2 \langle e_{\rmn g}^2
\rangle} \exp \left[\frac{\pi}{4} \left(\frac{e}{\langle e_{\rmn g} \rangle 
}\right)^2 \right],
\end{equation}
where the mean ellipticity is given by
\begin{equation}
\langle e_{\rmn g} \rangle = 
0.27 - \frac{m - 20}{25} 
\end{equation}
in the range of interest ($15 \la m \la 23$).
This distribution is shown in \fig{e_theory} for a number of 
different magnitudes.
For highly elliptical images the analytical function is too broad
-- $e > 1$ is permitted -- but the behaviour of ${\rmn d}p_{\rmn g}/{\rmn d}e$
is unimportant for $e \ga 0.5$, as galaxies are the only
objects with such high ellipticities.

\subsubsection{Galaxy redshift surveys}

A generic \grs\ catalogue is assumed to consist of all
images (in a certain region of the
sky) which are brighter than the magnitude limit, $m_{\rmn lim}$,
and have ellipticities greater than $e_{\rmn min}$.
For each source population the number over the whole sky is 
$N(<m_{\rmn lim}) [1 - p(< e_{\rmn min})|_{< m_{\rmn lim}}]$,
which must then be normalised by $N_{\rmn tot}$, the
number of objects in the survey.
The limitations of this model are discussed where relevant,
but, despite its simplicity, it provides a reasonable estimate
of the relative numbers of galaxies, stars and other objects
that are observed spectroscopically in a galaxy survey.
The ellipticity cut-off determines both the completeness, $C$, and 
efficiency, $E$, of a \grs\ of a given magnitude limit; the
resultant values of these parameters can then be 
used to assess the limitations of the model.

\begin{figure}
\includegraphics{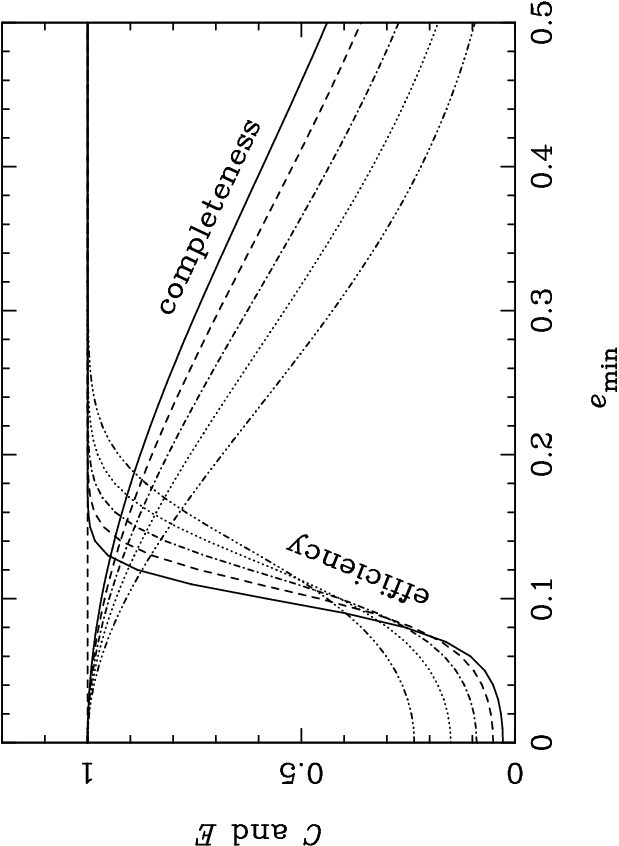}
\vspace{\singlefigureheight}
\caption{The completeness, $C$, and the efficiency, $E$, of a
\grs, as defined in \eqs{com} and (\ref{equation:eff}), 
shown as a function of the ellipticity cut-off, $e_{\rmn min}$.
Both are shown for five different limiting magnitudes:
$m_{\rmn lim} = 17$ (solid lines);
$m_{\rmn lim} = 18$ (dashed lines);
$m_{\rmn lim} = 19$ (dot-dashed lines);
$m_{\rmn lim} = 20$ (dotted lines);
and
$m_{\rmn lim} = 21$ (dot-dot-dot-dashed lines). 
The galaxy number counts are calculated assuming that
passive luminosity evolution cancels the $K$-corrections,
and an Einstein-de Sitter cosmological model is used.}
\label{figure:c and e}
\end{figure}

A survey's completeness is the fraction of all galaxies which 
appear in the sample, and is simply
\begin{equation}
\label{equation:com}
C(e_{\rmn min}, m_{\rmn lim}) 
= 1 - p_{\rmn g} (< e_{\rmn min})|_{<m_{\rmn lim}},
\end{equation}
where $p_{\rmn g} (< e)|_{<m}$ is given in \eq{dp_gde}.
Hence a survey is $100$ per cent complete if $e_{\rmn min} = 0$,
but becomes significantly incomplete for $e_{\rmn min} \ga 0.2$,
as shown in \fig{c and e}.

The efficiency of a galaxy survey is defined as the fraction of objects in 
the survey which actually are galaxies, and so is dependent 
also on the stellar population.
Using the stellar number counts and ellipticity distribution
of \eqs{n_stars} 
and (\ref{equation:dpde_stars}), respectively,
it is given by
\begin{equation}
\label{equation:eff}
E(e_{\rmn min}, m_{\rmn lim}) = 
\end{equation}
\[
\mbox{} \frac{1 - p_{\rmn g} (< e_{\rmn min})|_{<m_{\rmn lim}}}
{1 - p_{\rmn g} (< e_{\rmn min})|_{<m_{\rmn lim}}
+ \left[1 - p_{\rmn s} (< e_{\rmn min})|_{<m_{\rmn lim}} \right]
\frac{N_{\rmn s} (< m_{\rmn lim})}{N_{\rmn g} (< m_{\rmn lim})}}
,
\]
where $N_{\rmn g}(<m)$ is given in \eq{n_g_m}.
This is shown for surveys of varying depth in \fig{c and e},
and the limitations of ellipticity as a means of star-galaxy 
separation are exposed. For a given magnitude limit the
optimal choice of $e_{\rmn min}$ is close to 
the point where $C = E$, although 
slightly higher
completeness may be preferable.
Using $e_{\rmn min} \simeq 0.15$ can result in completenesses and 
efficiencies of $\sim 90$ per cent for $m_{\rmn lim} \simeq 19$, but
more sophisticated techniques based on surface brightness
profiles (\eg\ Maddox \etal\ 1990a) can achieve $\ga 95$ per cent
for both to fainter limits.

\begin{figure*}
\includegraphics{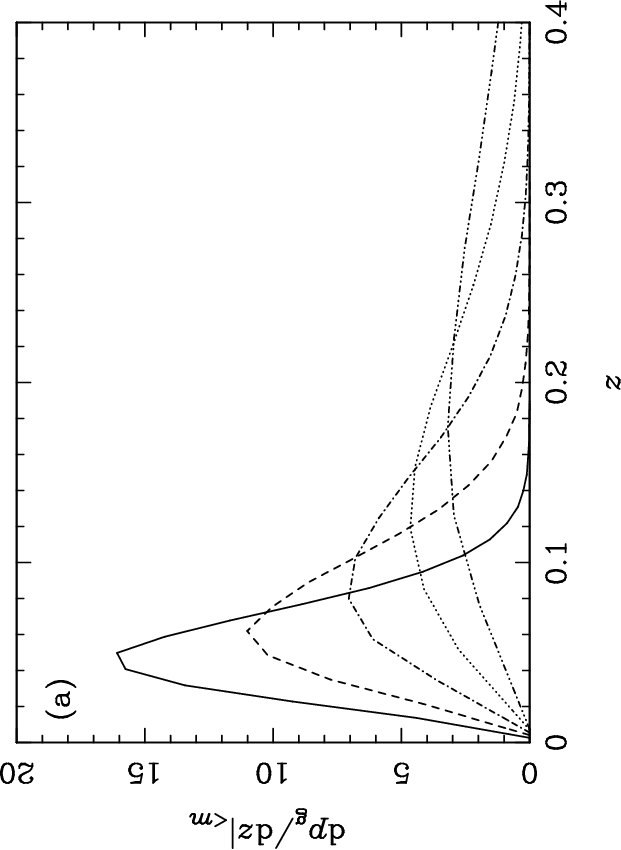}
\includegraphics{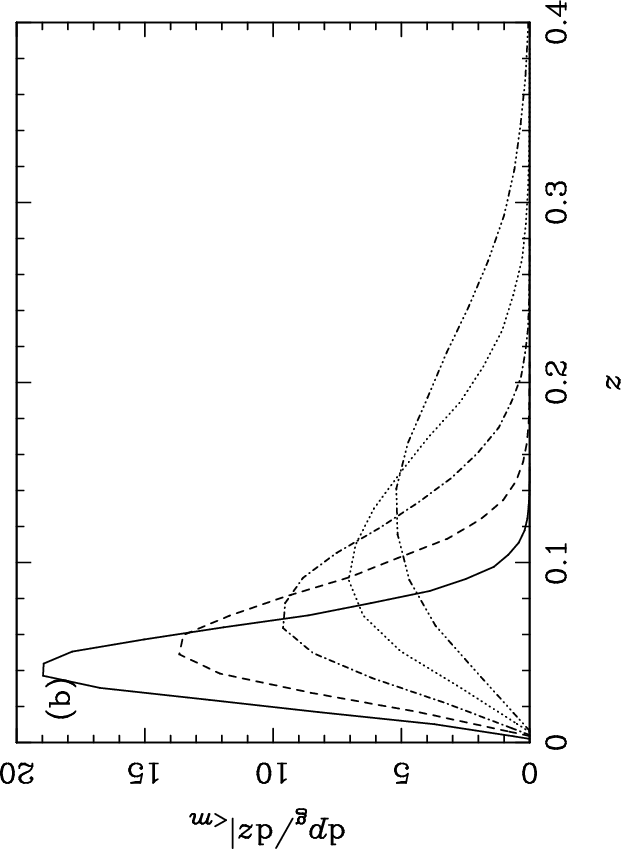}
\vspace{\singlefigureheight}
\caption{The redshift distribution of \grss\ of varying magnitude
limits:
$m_{\rmn lim} = 17$ (solid lines);
$m_{\rmn lim} = 18$ (dashed lines);
$m_{\rmn lim} = 19$ (dot-dashed lines);
$m_{\rmn lim} = 20$ (dotted lines);
and
$m_{\rmn lim} = 21$ (dot-dot-dot-dashed lines).
In (a) the galaxies' $K$-corrections are 
cancelled by passive luminosity evolution, whilst
in (a) there is no luminosity evolution, 
and 
the $K$-corrections
of Kinney \etal\ (1996) have been used.
An $\omo = 1$, $\olo = 0$ cosmology is assumed in all cases as
the cosmological model is unimportant at these depths.}
\label{figure:dpdz_m_app}
\end{figure*}

The redshift coverage of a given \grs\ is 
of intrinsic interest, and it can also determine the 
lensing statistics within it (\eg\ Kochanek 1992),
depending on the selection effects. 
The redshift distribution of the galaxies is not 
independent of the ellipticity cut-off -- the 
fainter and hence more distant galaxies are those 
likely to be omitted, reducing the high-redshift tail
of the sample -- but it is not a dominant effect, and 
is ignored here. 
From \eqs{schechter_grs} and (\ref{equation:dndmdz_gal}),
the a priori probability that a galaxy of magnitude $m$
has redshift $z$ is 
\begin{equation}
\label{equation:dpdz_m_app}
\left.\frac{{\rmn d}p_{\rmn g}}{{\rmn d}z}\right|_m = 
\frac{\frac{{\rmn d}V_0}{{\rmn d}z}
\left. \frac{{\rmn d}n_{\rmn g}}{{\rmn d}M} \right|_{M = M(m, z)}}
{\int_0^\infty \frac{{\rmn d}V_0}{{\rmn d}z^\prime}
\left. \frac{{\rmn d}n_{\rmn g}}{{\rmn d}M} \right|_{M = M(m, z^\prime)}
\, {\rmn d}z^\prime}.
\end{equation}
A number of fitting formul\ae\ have been developed 
to approximate this function 
(\eg\ Baugh \& Efstathiou 1993; Brainerd, Blandford \& Smail 1996), 
but the numerical 
integration is used here in order to look at the variation
with cosmological model and evolution.
The redshift distribution for the survey is then obtained 
by integrating \eq{dpdz_m_app} over $m$ and weighting
by the differential number counts [\eq{dndmdz_gal}],
which gives
\begin{equation}
\label{equation:dpdz_grs}
\left.\frac{{\rmn d}p_{\rmn g}}{{\rmn d}z}\right|_{<m} 
=
\frac{\int_{- \infty}^{m} \frac{{\rmn d}V_0}{{\rmn d}z}
\left. \frac{{\rmn d}n_{\rmn g}}{{\rmn d}M} \right|_{M = M(m^\prime, z)}
\, {\rmn d}m^\prime}
{\int_0^\infty \int_{- \infty}^{m}
\frac{{\rmn d}V_0}{{\rmn d}z^\prime}
\left. \frac{{\rmn d}n_{\rmn g}}{{\rmn d}M} \right|_{M = M(m^\prime, z^\prime)}
\,{\rmn d}m^\prime \, {\rmn d}z^\prime}
.
\end{equation}
This is shown in \fig{dpdz_m_app} for 
both the $E(z) = - K(z)$ model in (a) and the no-evolution model in (b).
The mean redshift of the 
surveys increase with $m_{\rmn lim}$, but the influence of 
the $K$-corrections becomes increasingly dominant as the surveys
go deeper. For $m_{\rmn lim} \simeq 20$, the 
difference between the two models 
is approximately equivalent to a changing $m_{\rmn lim}$ by one magnitude.

\subsection{Unlensed quasars}
\label{section:quasars}

The number counts of quasars brighter than the survey
limit are important in this calculation as those 
with high ellipticities can contaminate any sample of 
candidate lenses.
The faint end of the
quasar luminosity function is 
important too, as it affects lensing statistics
due to magnification bias. The redshift
distribution of the sources is also necessary, as lensing
probability increases so rapidly with source redshift
(\eg\ Turner, Ostriker \& Gott 1984; Kochanek 1993a).

\subsubsection{Number counts}

The differential quasar number counts,
as a function of magnitude, $m$, and
redshift, $z$, are given by
\begin{equation}
\label{equation:d2qdmdz}
\frac{{\rmn d}^2 N_{\rmn q}}{{\rmn d} m {\rmn d}z}
= \frac{3 [1 - (z / z_{\rmn max})^2]} {6 + z_{\rmn max}}
\frac{N_{\rmn q0}}
{10^{\alpha_{\rmn q}(m - m_{\rmn q0})}
     - 10^{\beta_{\rmn q}(m - m_{\rmn q0})}},
\end{equation}
where
$m_{\rmn{0}} = 19.0 \pm 0.2$ is the quasar break magnitude,
$\alpha_{\rmn q} = 0.9 \pm 0.1$
is the bright-end slope, $\beta_{\rmn q} = 0.3 \pm 0.1$
is the faint-end slope, and
$N_{\rmn{q}0} = (4.1 \pm 0.4) \times 10^5 = 10 \pm 1 \persqdeg$
(Boyle \etal\ 1988; Kochanek 1996a),
and $z_{\rmn max} = 3.2$ is chosen to match 
the redshift dependence of
the Large Bright Quasar Survey (Hewett \etal\ 1995).
Integrating
\eq{d2qdmdz}
over redshift and magnitude gives the
cumulative number counts as
\begin{equation}
\label{equation:n_q_m}
N_{\rmn q} (< m) =
\int_{- \infty}^m \frac{N_{\rmn q0}}
{10^{\alpha_{\rmn q}(m - m_{\rmn q0})}
- 10^{\beta_{\rmn q}(m - m_{\rmn q0})}}
\, {\rmn d}m ,
\end{equation}
which is plotted as the upper solid line in
\fig{num counts}.

The quasar and lens numbers calculated from these expressions
represent lower bounds on the true populations.
Firstly, most quasar surveys are based on prior colour-selection
(\eg\ the UV-excess
criteria of the Boyle \etal\ 1988 sample),
whereas any quasars found in redshift survey spectra are not 
subject to such cuts.
There is also some uncertainty in the number of
faint ($m \ga 21$) quasars, as 
the deepest quasar samples may be significantly incomplete.
This possibility can be tested by `complete'
spectroscopic surveys, in which there is no morphological or 
chromatic pre-selection.
Results from the deepest such sample,
the Fornax spectroscopic
survey (Drinkwater \etal\ 1999), are shown in \fig{num counts},
but they do not differ greatly from the 
Hartwick \& Schade (1990) compilation of data.
Clearly a complete spectroscopic survey to $m \ga 21$ would be of
great interest.

\subsubsection{Ellipticity distribution}

As with stars, quasars can appear extended on photographic plates
due to random noise. Their ellipticity distribution is likely
to be similar to that of stars, but with more outliers, due to
gravitational lensing and close quasar-galaxy associations. 
As strongly-lensed quasars are considered
explicitly here, the ellipticity distribution of unlensed
quasars is assumed to be the same as that of stars,
so that ${\rmn d}p_{\rmn q}/{\rmn d} e|_{<m}
= {\rmn d}p_{\rmn s}/{\rmn d} e|_{<m}$
[\eq{dpde_stars}
and \fig{e_theory}].

\subsection{Lensed quasars}
\label{section:lensed quasars}

The statistics of quasar lensing are most often framed in terms 
of the probability, $p_{\rmn q}$, that a given quasar is lensed.
Here, however, it is more useful to consider lensed quasars 
as a population of objects in their own right, distinct in
particular from the unlensed quasar population. 
The differential number of lenses with source redshift $z$ 
and magnitude $m$ is related to the lensing probability by
\begin{equation}
\frac{{\rmn d}^2 N_{\rmn l}}{{\rmn d}m {\rmn d} z}
= p_{\rmn q}  \,
\frac{{\rmn d}^2 N_{\rmn q}}{{\rmn d}m {\rmn d} z},
\end{equation}
where ${\rmn d}^2 N_{\rmn q} / {\rmn d}m {\rmn d} z$,
is given in 
\sect{quasars}.
The integral number counts of lenses is then simply
\begin{equation}
N_{\rmn l} (< m) = \int_{-\infty}^{m}
\int_0^\infty
p_{\rmn q}  \,
\frac{{\rmn d}^2 N_{\rmn q}}{{\rmn d}m^\prime {\rmn d} z}
\, {\rmn d}z
\,{\rmn d}m^\prime.
\end{equation}

The first step in the calculation of 
$p_{\rmn q}$ is to find, $p_{\rmn q,g}$,
the probability that the quasar (of magnitude $m$ and 
redshift $z$) is lensed by a given galaxy.
The lenses
that are missed in follow-up observations (those with
small image separations or high flux ratios) also have 
small ellipticities, and so are excluded from the model 
redshift surveys by the ellipticity
cut described below and in \sect{num counts}.

All three galaxy types are modelled as singular isothermal spheres
(Turner \etal\ 1984; Binney \& Tremaine 1987; 
Kochanek 1994), which are consistent with lens statistics
(\eg\ Kochanek 1993a, 1996a) but cannot produce the more complex 
image configurations observed.
Their mass distribution is characterised by
$\sigma$, the dark matter velocity dispersion,
which is marginally larger than the observed line-of-sight velocity dispersion,
$\sigma_{||}$, as shown by Kochanek (1994) and Mortlock \& Webster (2000b).
The line-of-sight dispersion is related to $M$,
the absolute magnitude of the lens galaxy,
by the Faber-Jackson (1976)
relationship for Es and S0s, and the Tully-Fisher (1977)
relationship\footnote{The Tully-Fisher (1977)
relationship actually gives the
circular speed, which is equal to $\sqrt{2} \sigma$.} for spirals.
Both conversions have the same mathematical form, so
\begin{equation}
\sigma \simeq 1.1 \sigma_{||}
= \sigma_* 10^{- 2/5 \, \gamma \, [M(m_{\rmn g}, z_{\rmn g}) - M_*]}
\end{equation}
where
$\gamma_{\rmn{E}} = \gamma_{\rmn{S0}} = 3.7 \pm 1$
and $\gamma_{\rmn{S}} = 2.6 \pm 1$
(de Vaucouleurs \& Olson 1982) and
$\sigma_* = 225 \pm 20$ km s$^{-1}$
for E-type galaxies, $\sigma_* = 205 \pm 20$ km s$^{-1}$ for
S0 galaxies and $\sigma_* = 143 \pm 10$ km s$^{-1}$ for spirals
(Efstathiou \etal\ 1988).

Ignoring any selection effects for the moment, $p_{\rmn q,g}$
is given by the standard expression 
(\eg\ Kochanek 1995; Mortlock \& Webster 2000b)
\begin{equation}
\label{equation:p_qg_grs}
p_{\rmn q,g} = 
\frac{\theta_{\rmn E}^2}{4}
\frac{\int_2^\infty \frac{8}{\mu_{\rmn tot}^3}
\left.\frac{{\rmn d}^2 N_{\rmn q}}{{\rmn d}z {\rmn d}m^\prime}\right|
_{m^\prime = m + 5/2 \log(\mu_{\rmn tot})} {\rmn d} \mu_{\rmn tot}}
{\frac{{\rmn d}^2 N_{\rmn q}}{{\rmn d}z {\rmn d}m}},
\end{equation}
where 
\begin{equation}
\label{equation:th_e_grs}
\theta_{\rmn E} = 4 \pi \left(\frac{\sigma}{c}\right)^2
\frac{d_{\rmn A}(z_{\rmn g}, z)}{d_{\rmn A}(0, z)}
\end{equation}
is the Einstein angle of the lens.
Here $d_{\rmn A}(0, z)$ and $d_{\rmn A}(z_{\rmn g}, z)$
are the angular diameter distances
from the observer to the quasar
and the galaxy to the quasar, respectively
(\eg\ Carroll \etal\ 1992; Schneider, Ehlers \& Falco 1992).

Integrating $p_{\rmn q,g}$ over the galaxy population
gives 
\label{equation:p_q_grs}
\begin{equation}
p_{\rmn q} = \int_0^{z} \int_{- \infty}^\infty
\frac{{\rmn d} V_0}{{\rmn d} z_{\rmn g}}
\left.\frac{{\rmn d} n_{\rmn g}}{{\rmn d} M}
\right|_{M = M(m_{\rmn g}, z_{\rmn g})}
p_{\rmn q,g} \,
{\rmn d}m_{\rmn g} \, {\rmn d}z_{\rmn g}.
\end{equation}
The choice of integration variables is arbitrary, but
$z_{\rmn g}$ and $m_{\rmn g}$ are used here as they are
observables. The resultant lens number counts are
shown for three different cosmological models 
in \fig{num counts}, but not all of these lenses are
detectable in practice due to various observational limitations.

\subsubsection{Ellipticity distribution}

The vast majority of lensed quasars are effectively a small
collection of stellar images, and whether they appear resolved,
extended, or point-like is strongly dependent on the \psf.
The exception to this is when the
lens galaxy is of comparable brightness to the quasar images
(\eg\ Q~2237+0305),
in which case the source-deflector composite almost certainly 
appears as an extended source; this case is dealt with separately.

The singular isothermal sphere lens model,
used above to calculate the number of lenses, 
is too simple to reproduce the observed variety of 
image configurations.
A quadrupole term
-- either intrinsic ellipticity of the lens or external shear,
as used here --
is required to generate the four- (and five-) image 
lenses observed.
The possibility of a small
core radius is less important, as in most cases it only results in 
an (additional) demagnified central image.
In the presence of a dimensionless external shear\footnote{The use of 
an elliptical lens model is more realistic, but more 
expensive computationally.
A shear of magnitude $\gamma_0$ is 
roughly equivalent to an ellipticity of $3 \gamma_0$ 
(\eg\ Keeton \& Kochanek 1996).} $\gamma_0$,
but with no core radius, the lens equation is
(\eg\ Kovner 1987; Schneider \etal\ 1992)
\begin{equation}
\label{equation:dimensionless lens equation_grs}
\left[ \! \! \begin{array}{c}
\beta_x \\ \\
\beta_y
\end{array} \!\! \right] =
\left[ \! \begin{array}{c}
\left(1 - \gamma_0 - \frac{\theta_{\rmn E}}{\theta}
\right) \theta_x \\ \\
\left(1 + \gamma_0 - \frac{\theta_{\rmn E}}{\theta}
\right) \theta_y 
\end{array} \! \right],
\end{equation}
where $\vect{\beta} = (\beta_x, \beta_y)$ 
is the angular source position,
$\vect{\theta} = (\theta_x, \theta_y)$ is the angular image position.
This can be solved using the method outlined by Schneider \etal\ (1992),
in which the radial components of the image positions
are given by the roots of a one-dimensional equation.

In order to obtain the ellipticity distribution of lensed quasars,
Monte Carlo simulations of 
$\sim 10^5$ lenses were generated for several different deflector models.
The integration over the quasar population was performed by
using $p_{\rmn q}$ [\eq{p_q_grs}] and ${\rmn d}^2 N_{\rmn q}/
{\rmn d}z {\rmn d}m$ to draw sources randomly from
population of lensed quasars, and a lens galaxy 
was similarly-generated for each source.
The source position, $\vect{\beta}$, was biased in favour of 
close alignments using the magnification bias described
above, but no other selection effects were included.
For each source-lens pair the lens equation was solved to
give image positions, $\vect{\theta}_i$ and magnifications,
$\mu_i$, with $1 \leq i \leq N_i$, where $N_i$ is the number 
of images (either two or four in this model).

The appearance of a given lens in the survey data 
is determined by the \psf, denoted by
$f_{\rmn see} (\vect{\theta}) = f_{\rmn see} (\theta)$,
which is 
normalised so that $\twoint f_{\rmn see} (\theta) \, {\rmn d}
\theta^2 = 1$.
The contribution from the galaxy is usually negligible
(See below.), and surface brightness-related selection effects are also
ignored here (but treated explicitly in Mortlock \& Webster 2000c).
The surface brightness
of a lens is then
\begin{equation}
f(\vect{\theta}) = F \frac{\sum_{i = 1}^{N_i} \mu_i f_{\rmn see}
\left( | \vect{\theta} - \vect{\theta}_i | \right)}
{\sum_{i = 1}^{N_i} \mu_i},
\end{equation}
where $F$ is the total observed flux of the quasar images.
From this the moments and ellipticity [as defined in \eq{e}]
can be found.

\begin{figure}
\includegraphics{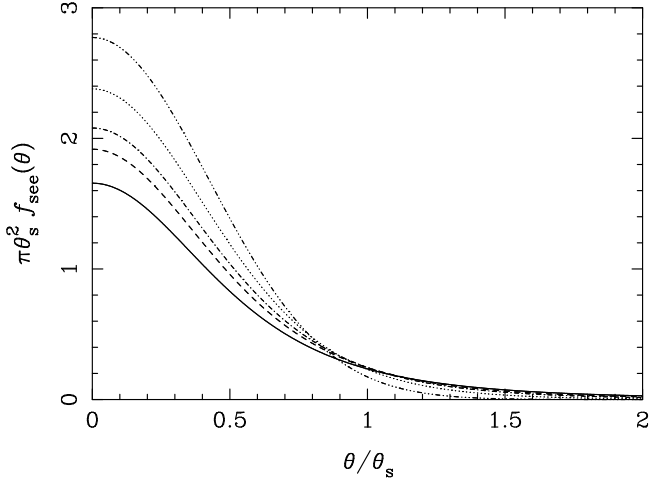}
\vspace{\singlefigureheight}
\caption{Dimensionless forms of the
\psf, $f_{\rmn see} (\theta)$,
under the assumption of a Moffat (1969) profile.
The profile is shown for different 
values of the $\eta$-parameter used in the definition
of the \psf\ [\eq{moffat}]:
$\eta = 2.0$ (solid line);
$\eta = 2.5$ (dashed line);
$\eta = 3.0$ (dot-dashed line);
$\eta = 5.0$ (dotted line);
and
$\eta \rightarrow \infty$ (\ie\ Gaussian seeing; dot-dot-dot-dashed line).}
\label{figure:seeing}
\end{figure}

In a similar Monte Carlo simulation,
Kochanek (1991) assumed Gaussian seeing,
ignoring the extended wings of 
the \psf\ (\eg\ King 1971).
A more general Moffat (1969) profile is used here,
for which the \psf\ is
\begin{equation}
\label{equation:moffat}
f_{\rmn see}({\theta}) =
\frac{(\eta - 1) (2^{1/\eta} - 1)}
{\pi (\theta_{\rmn{s}} / 2)^2}
\left[1 + (2^{1/\eta} - 1)
\frac{\theta^2}
{(\theta_{\rmn{s}} / 2)^2}\right]^{-\eta},
\end{equation}
which can be integrated to give
\begin{equation}
\label{equation:moffat_integrated}
F_{\rmn see} (< \theta) = 
1 - \left[1 + (2^{1/\eta} - 1)
\frac{\theta^2}
{(\theta_{\rmn{s}} / 2)^2} \right]^{1 - \eta},
\end{equation}
where $\theta_{\rmn{s}}$ is the full width at half-maximum
of the seeing disc
and 
$\eta$ determines its shape, 
as shown in \fig{seeing}. 
The profile cannot be normalised
if $\eta \leq 1$, and in the limiting case of $\eta \rightarrow \infty$
it becomes a Gaussian. 
The default
value assumed by the 
{\sc image reduction and analysis facility}
software
(Tody 1986) is $\eta = 2.5$,
but results such as those of Saglia \etal\ (1993)
imply that $\eta \simeq 3$. 
Conversely the areal profiles of stellar images in
the \APM\ survey shown in Maddox \etal\ (1990a) are close 
to log-linear, suggesting $\eta$ is very high.

\begin{figure*}
\includegraphics{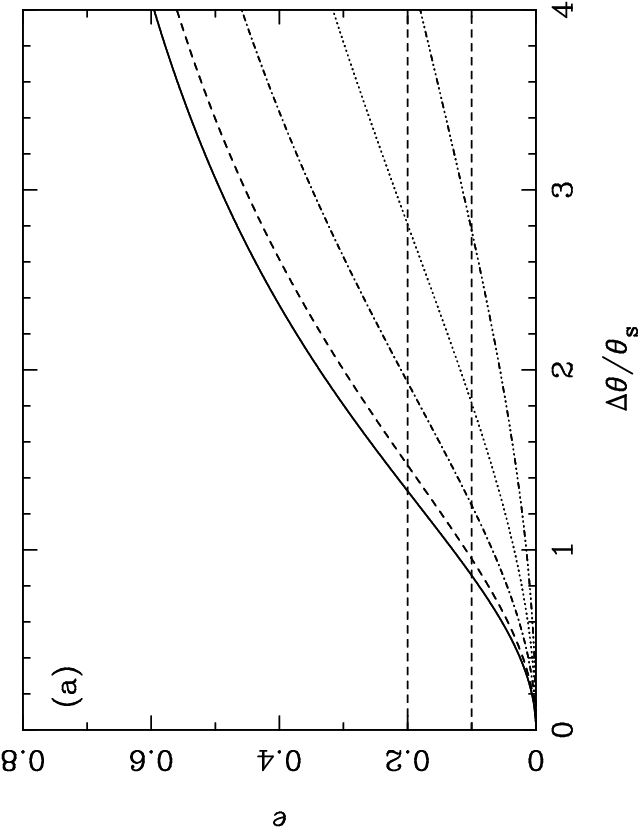}
\includegraphics{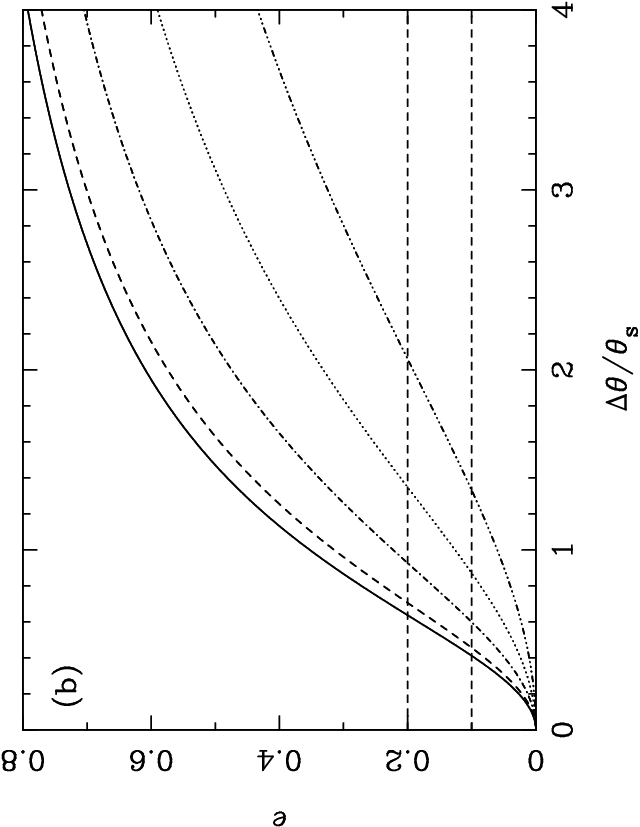}
\vspace{\singlefigureheight}
\caption{The ellipticity, $e$, of an image pair 
with angular separation $\Delta \theta$ (with
$\theta_{\rmn s}$ being the seeing).
Results are shown for various values of the magnitude difference between
the images: 
$\Delta m = 0$ (solid lines);
$\Delta m = 1$ (dashed lines);
$\Delta m = 2$ (dot-dashed lines);
$\Delta m = 3$ (dotted lines);
and
$\Delta m = 4$ (dot-dot-dot-dashed lines).
A Moffat (1969) \psf\ is assumed, with 
$\eta = 2.5$ in (a) and
$\eta \rightarrow \infty$ (\ie\ Gaussian seeing) in (b).
The horizontal dashed lines are at $e = 0.1$ and $e = 0.2$, reasonable
limits for the ellipticity cut-off used to select objects 
for galaxy surveys.}
\label{figure:e_pair}
\end{figure*}

Using this \psf\ the ellipticity can be found analytically, 
although the result is rather cumbersome, except if $N_i = 2$.
In this case it can be assumed,
without loss of generality, that both
images lie on the $x$-axis, 
and hence 
\eq{e} reduces to 
\begin{equation}
e = 1 - \left[ \frac{\langle \left(\theta_y - 
\langle \theta_y \rangle \right)^2 \rangle}
{\langle \left(\theta_x - 
\langle \theta_x \rangle \right)^2 \rangle} \right]^{1/2}.
\end{equation}
Using \eq{moffat}, this becomes
\begin{equation}
e = 
\end{equation}
\[
\mbox{} 1 - \left[
1 + 2 (\eta - 2) (2^{1/\eta} - 1) \frac{10^{2 \Delta m / 5}}
{(1 + 10^{2 \Delta m / 5})^2}
\left(\frac{\Delta \theta}{\theta_{\rmn s} / 2}\right)^2
\right]^{1/2},
\]
where $\Delta \theta$ is the angular separation of the images
and $\Delta m$ is their magnitude difference.
This is only defined for $\eta > 2$,
and reduces to the simpler 
Gaussian form found by Kochanek (1991) in the limit
of $\eta \rightarrow \infty$.
\Fig{e_pair} shows $e$ as a function of 
$\Delta \theta$ for several values of $\Delta m$ and 
two extreme values of $\eta$.
For a given image separation and seeing, 
$e$ decreases with increasing $\Delta m$, the pair becoming more like
a single point source. 
The ellipticity also decreases with decreasing $\eta$, 
the wings of the \psf\ `diluting' the 
ellipticity of the two images.
More generally, it is clear that small separation 
($\Delta \theta \la 1$ arcsec) lenses stand little 
chance of being classified as anything but stars.

\begin{figure*}
\includegraphics{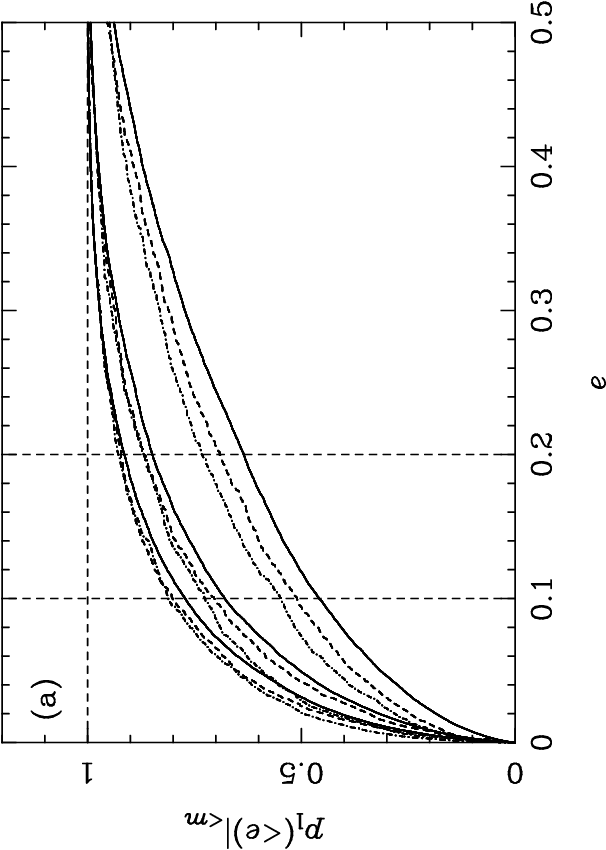}
\includegraphics{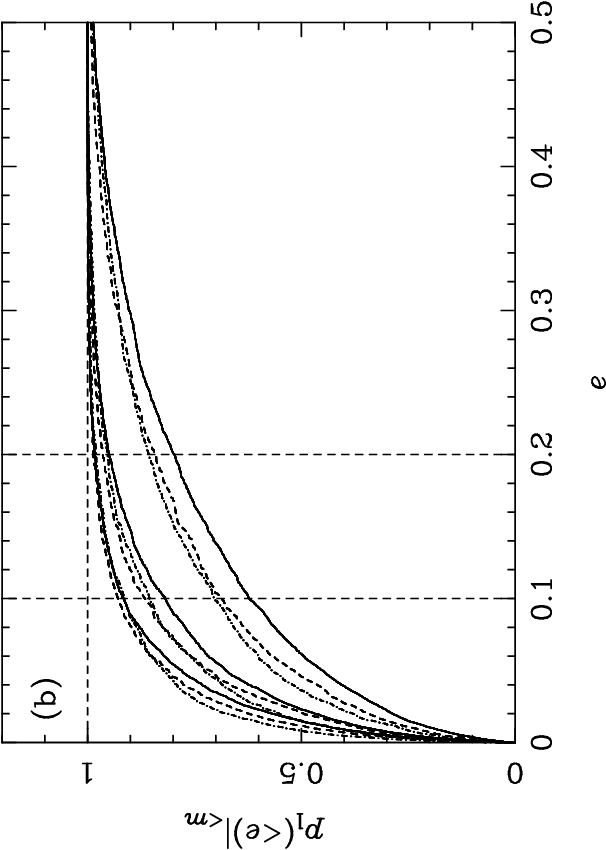}
\vspace{\singlefigureheight}
\caption{The cumulative distribution of ellipticities of lensed 
quasars, $p_{\rmn l} (< e)|_{<m}$, in seeing of 
$\theta_{\rmn s} = 2$ arcsec (a) and $\theta_{\rmn s} = 3$ arcsec (b).
In each case an Einstein-de Sitter cosmology is assumed and the magnitude 
limit is $m = 19$.
The lens model is a singular isothermal sphere with
an external shear of:
$\gamma_0 = 0.0$ (solid lines);
$\gamma_0 = 0.1$ (dashed lines);
and 
$\gamma_0 = 0.2$ (dot-dashed lines).
The three lines for each lens model are for different
values of the $\eta$-parameter used in the definition
of the \psf\ [\eq{moffat}]:
$\eta = 2.5$ (upper lines);
$\eta = 3$ (middle lines);
and
$\eta \rightarrow \infty$ (\ie\ Gaussian seeing; lower lines). 
The vertical dashed lines indicate reasonable values for
the ellipticity cut-off used to select objects for a
galaxy survey, and this can be used to obtain the fraction of 
lenses retained.}
\label{figure:p_e_lens}
\end{figure*}

Using the above formulation to calculate $e$ for each simulated 
lens gives a distribution, 
${\rmn d}p_{\rmn l}/{\rmn d}e|_{<m}$,
that is dependent mainly on $\gamma_0$ and
the \psf.
There is only weak dependence on the cosmological model,
the magnitude limit of the survey,
and the scale velocity dispersion of the galaxies 
(as $\langle \Delta \theta \rangle \propto \sigma_*^2$; 
Kochanek 1993b).
\Fig{p_e_lens} shows the cumulative ellipticity
distribution, $p_{\rmn l} (< e)|_{<m}$, for
various values of the model parameters.
The critical value is 
$1 - p_{\rmn l} (< e_{\rmn min})|_{<m}$, as this 
is the fraction of lenses that would enter any given \grs.
From the completeness and efficiency arguments 
in \sect{galaxies}, $0.1 \la e_{\rmn min} \la 0.2$;
this range is delineated by 
the vertical dashed lines.
An external shear of 0.2 
can reduce the number of 
`elliptical' (\ie\ $e > e_{\rmn min}$) lenses by
up to a factor of four, but the effects of the \psf\ are
clearly more important.
Increasing $\theta_{\rmn s}$ from
2 arcsec to 3 arcsec decreases the number of 
elliptical lenses by $\sim 30$ per cent in Gaussian seeing,
but results in an order of magnitude decrease if 
$\eta \simeq 2.5$. 
With $\eta \la 3$
the \psf\ is broader, and
$0.1 \la p_{\rmn l} (< e)|_{<m} \la 0.3$ (for 2 arcsec seeing)
or 
$0.05 \la p_{\rmn l} (< e)|_{<m} \la 0.2$ (for 3 arcsec seeing).
The strong dependence on the shape of the \psf\ shows that
the wings do contribute significantly to the ellipticity
of multiple images, contrary to the assumption 
of Kochanek (1991).
The variation in \psf\ between different sets of observational parameters
thus places limits on the accuracy of any 
generic predictions of the number of lenses 
that will enter a \grs.

\subsubsection{Lenses with visible deflectors}

In the most pessimistic scenario presented above, only
about 5 per cent of all lensed quasars are sufficiently elliptical 
to enter a \grs\ as candidate galaxies, in which case 
there would be only limited use in searching \grs\ spectra
for potential lenses.
However the simulation did not take into account the
light from the deflector.
If the lens galaxy is bright enough for the
source-deflector composite image to appear extended,
but the quasar is also bright enough that the composite
spectrum is distinguishable from that of an isolated galaxy,
the lens will both enter the redshift survey
and be detectable spectroscopically.
Kochanek (1992) performed an `initial, order of magnitude'
estimate of the statistics of such lenses, 
but 
the contribution of the quasar light to the 
observed magnitude of the compound object was 
neglected. This is an important effect, as it
increases the effective depth of the lens survey by $\sim 1$ mag,
and can result in up to an order of magnitude more lenses.

\begin{figure}
\includegraphics{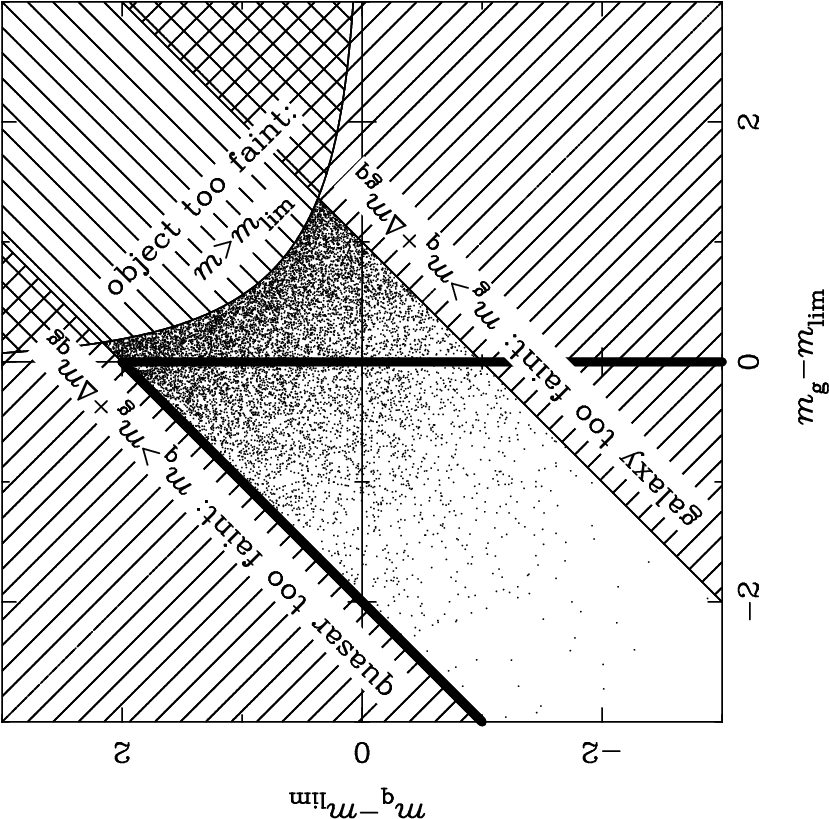}
\vspace{\squarefigureheight}
\caption{The range of deflector and source magnitudes ($m_{\rmn g}$ and 
$m_{\rmn q}$, respectively) probed by a \grs\ of magnitude limit
$m_{\rmn lim}$. 
Lenses with $- 5/2\, \log(10^{-2/5\, m_{\rmn q}} + 10^{-2/5\, m_{\rmn g}}) 
> m_{\rmn lim}$ are omitted as the 
combined flux from both the quasar and lens galaxy is below the
flux limit.
The survey is assumed to have 
$\Delta m_{\rmn qg} = 2$ (all lenses with $m_{\rmn q} - m_{\rmn g} >
\Delta m_{\rmn qg}$ being omitted as the quasar is too faint to
be detectable spectroscopically)
and $\Delta m_{\rmn gq} = 1$
(all lenses with $m_{\rmn g} - m_{\rmn q} >
\Delta m_{\rmn gq}$ being omitted as the galaxy is too faint to
make the lens appear extended).
Hence only those lenses which lie in the un-hatched region are 
included. The heavy solid lines show the boundaries of the equivalent
region used by Kochanek (1992) -- the 
inclusion of the lenses for which 
$m_{\rmn g} > m_{\rmn lim}$ is far more important than
exclusion of the lenses
in which the galaxy is too faint. This last point is illustrated
by the scatter plot within the allowed region, which shows 
the distribution of galaxy-quasar pairs generated from their 
respective number counts [\eqs{n_g_m} and (\ref{equation:n_q_m})]. 
For the scatter plot $m_{\rmn lim} = 19$ was assumed, and 
the quasars, being strongly lensed, are assumed to be 
magnified by a factor of 4 on average.}
\label{figure:delta_m_grs}
\end{figure}

Defining $m_{\rmn g}$ as the magnitude of (just) the galaxy,
and $m_{\rmn q}$ as the total, lensed magnitude of all the 
quasar images, the observed magnitude of a lens is  
given by 
\begin{equation}
\label{equation:m_conv}
m = - 5/2\, \log \left(10^{-2/5\, m_{\rmn q}} 
+ 10^{-2/5\, m_{\rmn g}} \right).
\end{equation}
As illustrated in \fig{delta_m_grs},
a lens with these properties will be included in a
redshift survey of limiting 
magnitude $m_{\rmn lim}$ 
if the following three 
criteria are met:

\begin{enumerate}

\item{The lens must be bright enough to enter the \grs\
	at all, \ie\ $m < m_{\rm lim}$, which is
	shown by the curved line in \fig{delta_m_grs}.
	This gives the extra depth relative to the redshift
	survey proper, for which the flux cut-off is
	$m_{\rmn g} < m_{\rmn lim}$.}

\item{The quasar must be sufficiently bright, relative
	to the lens galaxy, that its emission lines 
	are detectable in the \grs\ spectrum.
	Following Kochanek (1992), this requirement
	is modelled as
	$m_{\rmn g} - m_{\rmn q} < \Delta m_{\rmn qg}$,
	which is given by the upper of the two diagonal
	cuts in \fig{delta_m_grs}. Due to the strong,
	broad emission lines present in most quasar
	spectra (\eg\ Peterson 1997), $\Delta m_{\rmn qg} \simeq 2$
	is assumed unless stated otherwise. This 
	value is also supported by the discovery of 
	Q~2237+0305 as described in Kochanek (1992).}

\item{The galaxy must be sufficiently bright that the 
	the lens is classified as a galaxy in the low-resolution
	data from which the \grs\ is selected.
	This is assumed to be the case if
	$m_{\rmn q} - m_{\rmn g} < \Delta m_{\rmn gq}$,
	which is shown as the lower diagonal cut in 
	\fig{delta_m_grs}.
	The value of $\Delta m_{\rmn gq}$ is somewhat uncertain, 
	and could even be negative\footnote{If 
	$\Delta m_{\rmn gq} < - \Delta m_{\rmn qg}$, then 
	the only lenses included would
	be those with quasars too faint to be detectable, and the
	lensing probability would be zero.}
	if galaxies with superimposed stellar components 
	are reliably removed from the sample.
	In the case of Q~2237+0305, 
	$m_{\rmn q} \simeq m_{\rmn g} + 1$, 
	but the galaxy appears unremarkable at low resolution,
	suggesting that $\Delta m_{\rmn gq} \simeq 0$. 
	The simulations of the \APM\ galaxy survey 
	presented in Mortlock \& Webster (2000c) further imply that 
	$0 \la \Delta m_{\rmn gq} \la 1$.
	In other words, the galaxy need contribute only 
	half the flux of a lens for it to be classified 
	as non-stellar, and hence included in a \grs.
	If the `elliptical' lenses considered previously
	enter \grss, they could be included in
	this formalism by taking $\Delta m_{\rmn gq} 
	\rightarrow \infty$.}

\end{enumerate}

It is now possible to calculate 
$p_{\rmn q,g}$, the probability that a given galaxy (with
redshift $z_{\rmn g}$ and
magnitude $m_{\rmn g}$)
lenses a given quasar [with redshift $z$,
and lensed magnitude $m_{\rmn q}$, defined in terms of $m$ and $m_{\rmn g}$,
in \eq{m_conv}].
Adjusting \eq{p_qg_grs}, the lensing probability becomes
\begin{equation}
p_{\rmn q,g} = 
\end{equation}
\[
\frac{\theta_{\rmn E}^2}{4} 
\int_{2}^{\infty} \frac{8}{\mu_{\rmn tot}^3} 
\left.\frac{{\rmn d}^2N_{\rmn q}}{{\rmn d}z {\rmn d}m^\prime}
\right|_{m^\prime = - 5 / 2 \, \log\left(
10^{-2/5 \, m} - 10^{-2/5 \, m_{\rmn g}}
\right)} 
{\rmn d}\mu_{\rmn tot}, 
\]
if $m_{\rmn g,min} \leq m_{\rmn g} \leq m_{\rmn g,max}$,
but is zero if $m_{\rmn g}$ is outside these limits.
These, in turn, are given by
\begin{equation}
m_{\rmn g,min} = m + \frac{5}{2}
\log\left( 1 + 10^{-2/5 \, \Delta m_{\rmn qg}}\right)
\end{equation}
and
\begin{equation}
m_{\rmn g,max} = m + \frac{5}{2}
\log\left( 1 + 10^{2/5 \, \Delta m_{\rmn gq}}\right).
\end{equation}
The integral is along a line of 
constant $m$, as shown in \fig{delta_m_grs}.
Integrating $p_{\rmn q,g}$ over the galaxy population yields
\begin{equation}
\frac{{\rmn d}^2N_{\rmn l}}{{\rmn d}m {\rmn d}z} =
\end{equation}
\[
\mbox{}
\int_0^{\rmn z_{\rmn q}} 
\int_{m_{\rmn g,min}}^{m_{\rmn g,max}}
\frac{{\rmn d}V_0}{{\rmn d}z_{\rmn g}} 
\left.\frac{{\rmn d}n_{\rmn g}}{{\rmn d}M}
\right|_{M = M(m_{\rmn g}, z_{\rmn g})} 
p_{\rmn q,g} \, {\rmn d}m_{\rmn g} \, {\rmn d}z_{\rmn g},
\]
where the integral over the deflector magnitude is restricted again
by the requirement that the overall magnitude of the lens be $m$.
Finally the cumulative number counts of such lenses are
\begin{equation}
N_{\rmn l} (< m) = \int_{-\infty}^{m} \int_0^{\infty}
\frac{{\rmn d}^2N_{\rmn l}}{{\rmn d}m {\rmn d}z}
\, {\rmn d}z.
\end{equation}
The values of $N_{\rmn l}$ implied are discussed in 
\sect{num_grs}, but it is immediately clear from the distribution of 
galaxy-quasar pairs in \fig{delta_m_grs} that 
the majority of the lenses are included only if both the quasar
and galaxy fluxes are explicitly accounted for in the calculation.

\section{Results}
\label{section:results}

Combining the various populations 
of objects considered in \sect{num counts}, it is possible
to make predictions about the number of lenses found
in particular redshift surveys (\sect{num_grs})
as well as the usefulness of this search
method (\sects{control}, \ref{section:e and c}
and \ref{section:z_dist_grs}). 

\subsection{Number of lenses}
\label{section:num_grs}

From the number counts 
given in \sect{num counts}, the number
of objects of type $t$
which enter a given redshift survey is 
\begin{equation}
N_t = N_{\rmn tot}
\frac{N_t(< m_{\rmn lim}) [1 - p_t(<e_{\rmn min})|_{< m_{\rmn lim}}]}
{\sum_{t^\prime} N_{t^\prime}(< m_{\rmn lim})
[1 - p_{t^\prime}(<e_{\rmn min})|_{< m_{\rmn lim}}]},
\end{equation}
where $N_{\rmn tot}$ is the total number of objects in the survey, 
$m_{\rmn lim}$ the magnitude limit
and 
$e_{\rm min}$ the ellipticity cut-off.
The sum is over all the populations
(\ie\ $t$ = s, g, q or l, for stars, galaxies, unlensed quasars
and lenses, respecitvely), although only
stars and galaxies make a significant contribution to the
denominator.

\begin{figure*}
\includegraphics{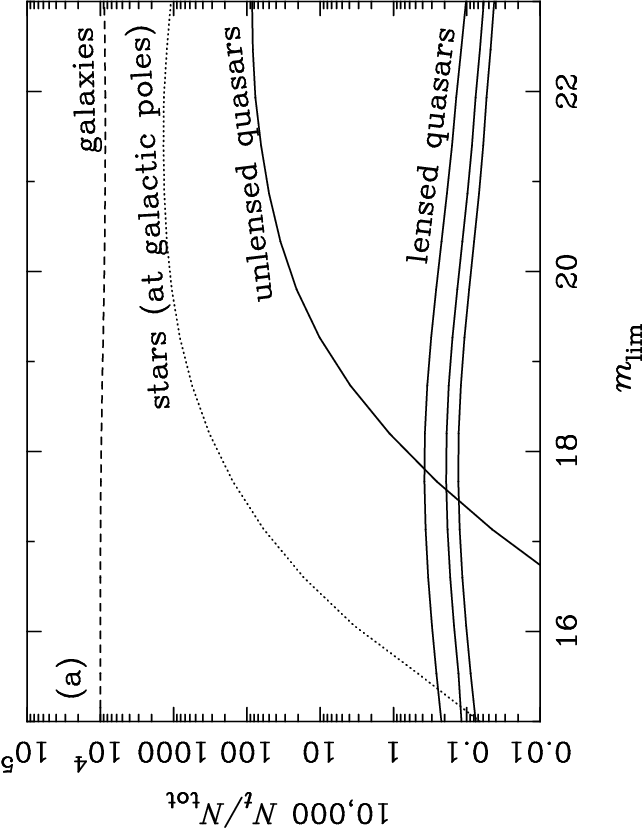}
\includegraphics{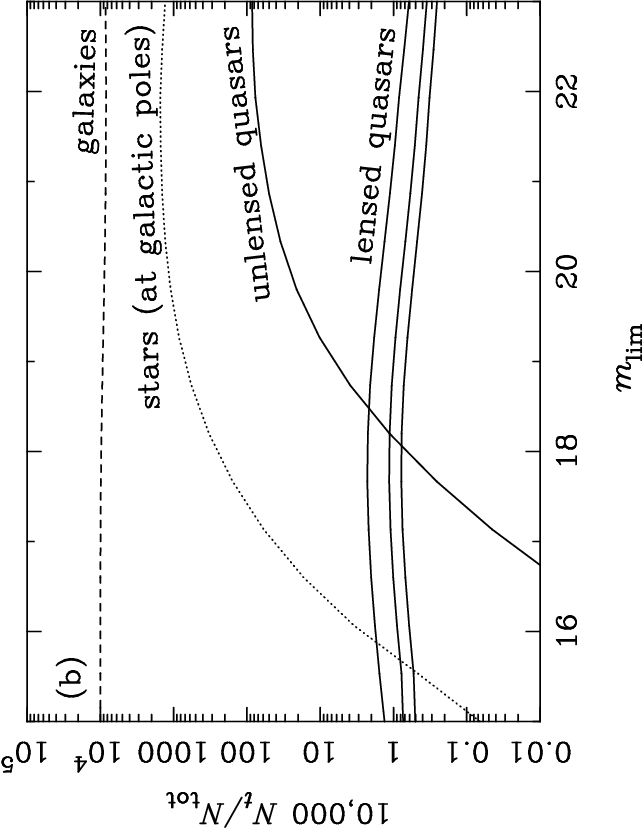}
\vspace{\singlefigureheight}
\caption{The expected number, $N_t$, of galaxies, stars, unlensed quasars
and lensed quasars
in a \grs\ as a function of the magnitude limit, $m_{\rmn lim}$.
The number of lenses is calculated using the spherical
lens model (\ie\ $\gamma_0 = 0$) in three cosmologies:
$\Omega_{\rmn m_0} = 1.0$ and $\Omega_{\Lambda_0} = 0.0$ (lower solid
line);
$\Omega_{\rmn m_0} = 0.3$ and $\Omega_{\Lambda_0} = 0.0$ (middle solid
line);
and 
$\Omega_{\rmn m_0} = 0.3$ and $\Omega_{\Lambda_0} = 0.7$ (upper solid
line).
The \grs\ is selected using an ellipticity cut $e_{\rmn min} = 0.15$,
and normalised to $N_{\rmn tot} = 10^4$.
The parent survey is assumed to have 
3 arcsec seeing
and a Moffat (1969) \psf\ with $\eta = 2.5$ in (a) and 
$\eta \rightarrow \infty$ (\ie\ Gaussian seeing) in (b).}
\label{figure:n_lens_grs}
\end{figure*}

\Fig{n_lens_grs} shows the number of `elliptical' lenses (and other objects)
per $10^4$ redshifts as a function of
$m_{\rmn lim}$ (chosen to facilitate comparison
with Kochanek 1992)
for a number of different observing conditions.
The lens fraction remains essentially constant
at about 1 in $10^5$ (a) or 1 in $10^4$ (b),
decreasing very slowly as $m_{\rmn lim}$ increases, due to
the decreased magnification bias.
For bright surveys this is far greater than the
numbers of lenses with bright deflectors considered
by Kochanek (1992), but it is not at all certain that
the `elliptical' lenses will enter a typical \grs.

\begin{figure*}
\includegraphics{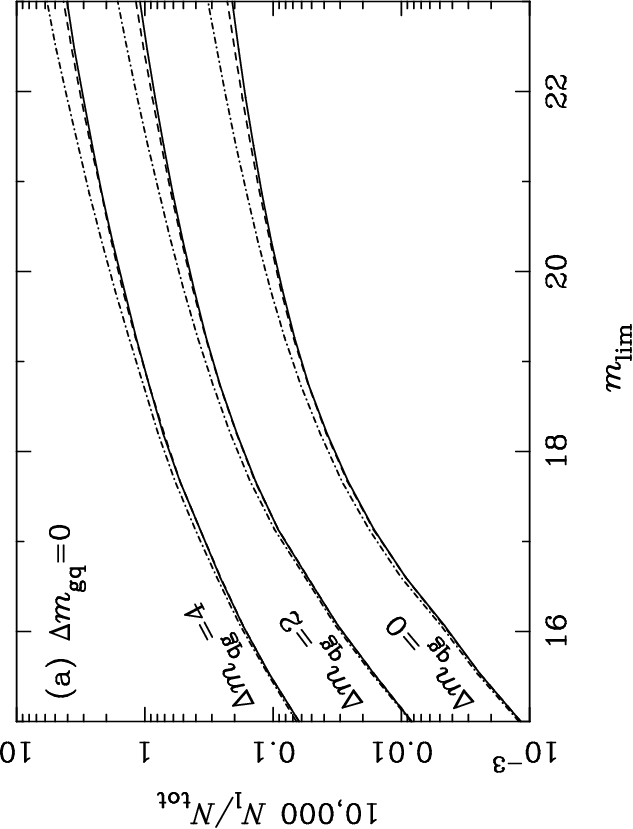}
\includegraphics{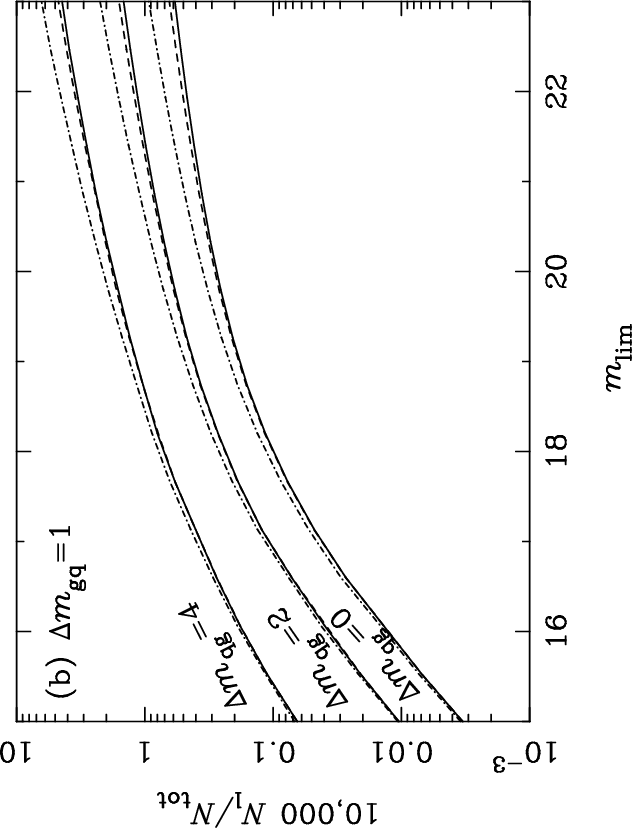}
\vspace{\singlefigureheight}
\caption{The expected number of lensed quasars, $N_{\rmn l}$,
per $10^4$ measured redshifts, shown (as in Kochanek 1992)
as a function of magnitude limit, $m_{\rmn lim}$.
Highly elliptical sources are not included in 
the survey; only galaxies and lenses with 
$- \Delta m_{\rmn gq} < m_{\rmn q} - m_{\rmn q} < \Delta m_{\rmn qg}$
and $- 5/2 \log(10^{-2/5 \, m_{\rmn q}} + 10^{-2/5 \, m_{\rmn g}}) <
m_{\rmn lim}$ are included.
(See \sect{lensed quasars} for a full explanation of 
these selection criteria.)
In each panel the three sets of lines represent different
values of the spectral prominence of quasars:
$\Delta m_{\rmn qg} = 0$ (lower lines);
$\Delta m_{\rmn qg} = 2$ (middle lines);
and 
$\Delta m_{\rmn qg} = 4$ (upper lines).
The morphological threshold for galaxy selection is 
assumed to be $\Delta m_{\rmn gq} = 0$ in (a) 
and $\Delta m_{\rmn gq} = 1$ in (b).
For each set of observational limits, three cosmological
models are shown:
$\Omega_{\rmn m_0} = 1.0$ and $\Omega_{\Lambda_0} = 0.0$ (solid lines);
$\Omega_{\rmn m_0} = 0.3$ and $\Omega_{\Lambda_0} = 0.0$ (dashed lines);
and
$\Omega_{\rmn m_0} = 0.3$ and $\Omega_{\Lambda_0} = 0.7$ (dot-dashed lines).}
\label{figure:n_lens_grs_min}
\end{figure*}

Multiply-imaged quasars with lens galaxies close to
the survey limit should definitely enter galaxy samples,
and so they give a lower bound on $N_{\rmn l}$.
As can be seen from \fig{n_lens_grs_min},
these lenses are considerably rarer than the `elliptical'
lenses at bright magnitudes, but their numbers increase
rapidly with $m_{\rmn lim}$, due to the steepness of
the quasar luminosity function (\sect{quasars}).
\Fig{n_lens_grs_min} can be compared directly with
the results of Kochanek (1992), although $N_{\rmn l}$
is higher by a factor of $\sim 5$,
due to the inclusion of the quasars' light in the
calculation.
This immediately implies that the majority of any such lensed
quasars found in redshift surveys should have lens
galaxies which are actually fainter than the survey limit,
as can be seen from \fig{delta_m_grs}.
Other than the depth of the survey,
$N_{\rmn l}$ is determined mainly by
$\Delta m_{\rmn qg}$, with $\Delta m_{\rmn gq}$
somewhat less important.
The strong dependence of $N_{\rmn l}$ on $\Delta m_{\rmn qg}$ 
is again purely a function of the quasar number counts,
but, being a measure of how sensitive a redshift survey is
to the presence of quasar emission features, it 
can be determined `experimentally' by running 
the survey software on the spectra of simulated lenses.
The value of $\Delta m_{\rmn gq}$ is more difficult to 
determine; fortunately it does not greatly affect the
lens statistics, provided that $\Delta m_{\rmn qg} \ga 1$.
The cosmological model
also has little impact on the number of lenses with
visible deflectors, as the lens galaxies are so close-by
that the observer-source and deflector-source
distances are approximately equal (Kochanek 1992).

These above results are summarised for a number of real
redshift surveys in
\tabl{n_lens_grs}.
None of the surveys have been systematically searched for lenses,
although all the 
\CfA\ survey spectra
were examined by eye, and it was process that led to the discovery
of 
Q~2237+0305 as a lens.
If `elliptical' lenses were included in the survey,
the detection of a lens is not unexpected;
however
the \CfA\ \grs\ is so bright ($\langle m \rangle \simeq 15$)
that multiple point sources should have been reasonably easy to remove
from the sample.
Assuming only those lenses with
visible, low-redshift deflectors were included,
the expected number of lenses
in the \CfA\ survey is $\sim 0.03$.
In other words, about 1 in 30 \CfA-like
galaxy samples should contain a spectroscopic quasar lens.
This is considerably higher
than the previous estimates of $\sim 0.0002$
(Huchra \etal\ 1985)
and $\sim 0.003$ (Kochanek \etal\ 1992);
the reason for this increase is simply the inclusion of
the quasars' light in the calculation.
Note also that Q~2237+0305 is actually fainter than the survey
limit, which is consistent with the conclusion that
galaxies with $m_{\rmn g} \ga m_{\rmn lim}$
dominate the lens statistics.

Of the other large redshift surveys that have already been completed,
both the European Southern Observatory Slice Project 
(\ESP; Vettolani \etal\ 1997) and 
the Canadian Network for Observational Cosmology (\CNOC 2)
sample (Yee, Ellingson \& Carlberg 1996) are too small to present
any real likelihood of lensing.
The million galaxies in the 
\MRP\ (Schuecker \etal\ 1996) could include a number 
of lenses, 
although only objective prism spectra 
were used for the redshift determination. 
The selection of candidate quasars from low-grade spectra
is an established technique
(\eg\ Hewett \etal\ 1995), 
but this would represent a considerable undertaking,
as the low resolution data would generate
a large number of false candidates lenses for which
further observations would be required.
Thus the most promising of the completed galaxy surveys is 
the \LCRS\ (Shectman \etal\ 1996) -- 
it is larger than the \CfA\ survey, as well as being deeper
($\langle m \rangle \simeq 19$), and should contain
one or two lenses (\cf\ \fig{n_lens_grs_min}).

The two surveys listed in \tabl{n_lens_grs} 
which have not been completed are also the 
most ambitious, and are likely to yield the most lenses.
The \TdF\ \grs\ (\eg\ Colless 1999; Folkes \etal\ 1999)
should contain at least 10 lenses,
and as many as 50 if the cosmological model and observational
parameters are optimal. Either way, it
would represent the largest sample
of lensed quasars generated from a single survey.
However the \TdF\ instrument has very small
optical fibres (effective radius 1 arcsec), 
and is unusually sensitive to a number of surface brightness-related
selection effects. The impact of these on $N_{\rmn l}$ is 
studied in more detail in Mortlock \& Webster (2000c).

The \SDSS\ (\eg\ Szalay 1998; Loveday \& Pier 1998)
is four times the size of \TdF,
and the spectra will be of considerably better quality.
Assuming $\Delta m_{\rmn qg} \simeq 4$ implies that
$N_{\rmn l} \ga 100$.
Further, the survey will also
include high-resolution imaging
of $\sim 10^4 \sqdeg$, which should allow the
morphological identification of an even larger number of lens
candidates.
The number counts shown in \fig{num counts}
then imply that the \SDSS\ should contain
between 200
(if $\Omega_{\rm m_0} = 1.0$ and $\Omega_{\Lambda_0} = 0.0$)
and 1000 
(if $\Omega_{\rm m_0} = 0.7$ and $\Omega_{\Lambda_0} = 0.3$).

\begin{table*}
\begin{minipage}{174mm}
\caption{The frequency of lensing in existing and planned \grss.}
\begin{tabular}{lrrrrrrrrr}
\hline 
\multicolumn{1}{c}{Survey}
&
\multicolumn{1}{c}{Reference}
&
\multicolumn{2}{c}{$m_{\rmn lim}$}
& 
\multicolumn{1}{c}{$N_{\rmn tot}$}
& 
\multicolumn{4}{c}{$N_{\rmn l}$}
&
\multicolumn{1}{c}{$N_{\rmn l,min}$}
\\
& & & & &
\multicolumn{2}{c}{{$\Omega_{\rmn m_0} = 1$}}
& 
\multicolumn{2}{c}{{$\Omega_{\rmn m_0} = 0.3$}}
&
\multicolumn{1}{c|}{{$\Omega_{\rmn m_0} = 1$}}
\\
& & & & &
\multicolumn{2}{c}{{$\Omega_{\Lambda_0} = 0$}}
& 
\multicolumn{2}{c}{{$\Omega_{\Lambda_0} = 0.7$}}
&
\multicolumn{1}{c|}{{$\Omega_{\Lambda_0} = 0$}}
\\
& & & & &
\multicolumn{1}{c}{$\eta = 2.5$}
& 
\multicolumn{1}{c}{$\eta = \infty$}
& 
\multicolumn{1}{c}{$\eta = 2.5$}
&
\multicolumn{1}{c}{$\eta = \infty$}
&
\\ & & & & & & & & & \\
\CNOC 2 & Yee \etal\ (1996) & $R = 21.5$ & $B_{\rmn J} = 22.8$
& $2.6 \times 10^3$
& 0.01 & 0.03 & 0.08 & 0.2 & 0.3
\\
\ESP\ & Vettolani \etal\ (1997) & $B_{\rmn J} = 19.4$
& $B_{\rmn J} = 19.4$ & $3.3 \times 10^3$
& 0.04 & 0.1 & 0.2 & 0.7 & 0.1
\\
\CfA\ & $\!\!\!\!\!\!$ Geller \& Huchra (1989) & $B_{\rmn Zwicky} = 15.5$ &
$B_{\rmn J} = 15.5$ & $1.5 \times 10^4$
& 0.1 & 0.4 & 0.8 & 2.3 & 0.03
\\
\LCRS\ & Shectman \etal\ (1996) & $R = 17.7$ & $ B_{\rmn J} = 19.2$ 
& $2.6 \times 10^4$
& 0.3 & 0.8 & 2.3 & 5.7 & 0.9
\\
\TdF\ & Colless (1999) & 
$B_{\rmn J} = 19.45\!\!\!$ & $B_{\rmn J} = 19.45\!\!\!$ & $2.5 \times 10^5$
& 2.5 & 7.5 & 18 & 50 & 10
\\
\MRP\ & Schuecker \etal\ (1996) & $J = 20$ & $ B_{\rmn J} = 19.5$
& $9.0 \times 10^5$
& 4.5 & 18 & 36 & 90 & 27
\\
\SDSS\ & Szalay (1998) & $r_{\rmn Gunn} = 18.2$ & $B_{\rmn J} = 18.9$ 
& $1.0 \times 10^6$
& 10 & 40 & 90 & 250 & 100
\\
\hline
\end{tabular}

\label{table:n_lens_grs}

A summary of the number of lenses expected 
in several completed and future redshift surveys. 
For each survey: $m_{\rmn lim}$ is the magnitude limit
in the band indicated (with approximate conversions 
to $B_{\rmn J}$, based on the redshift coverage of the survey); 
$N_{\rmn tot}$ is the total number of redshifts measured;
$N_{\rmn l}$ is the expected number of lenses;
and $N_{\rmn l,min}$ is the expected number of 
lenses if only those with visible detectors are 
included in the survey. 
The variation of $N_{\rmn l}$ with cosmological model 
and \psf\ is given explicitly, but the spherical lens model
and 3 arcsec seeing are assumed. 
On the other hand, 
$N_{\rmn l,min}$ is strongly dependent on neither the
cosmology, nor the \psf\ in the relevant magnitude range,
and so is only shown for the Einstein-de Sitter model.
In all cases $N_{\rmn l,min}$ is calculated assuming
$\Delta m_{\rmn gq} = 0$ and $\Delta m_{\rmn qg} = 2$,
except for the \MRP\ (which uses objective prism spectra,
and so has $\Delta m_{\rmn qg} \simeq 0$) 
and the \SDSS\ (which will obtain high-quality spectra with 
$\Delta m_{\rmn qg} \simeq 4$).
\end{minipage}
\end{table*}

\subsection{Control samples}
\label{section:control}

Kochanek (1992) suggested two
control samples -- foreground stars and unlensed quasars --
from which the efficiency and completeness of the
lens search could be calibrated.
However these objects are only useful controls if they enter
the \grs\ mainly through chance superpositions with survey galaxies.
Unfortunately, most of the stars and unlensed quasars in the survey
are 
not there due to chance alignments, but misclassification
(\eg\ \fig{n_lens_grs}).
For instance, the limitations \APM\ star-galaxy separation algorithm
(\eg\ Maddox \etal\ 1990a) imply that,
of the $\sim 2.5 \times 10^5$ objects in the \TdF\ \grs, 
several thousand will be stars, compared to only several
tens of star-galaxy superpositions.
With the exception of the brightest surveys
($m_{\rmn lim} \la 18$), the same is also true for unlensed quasars.

This effectively leaves this lens search technique without
any simple calibration, which is not important if the aim
is simply to discover new lenses, but must be addressed if
lensing probabilities are to be determined.
One possibility is to perform both
morphological and spectroscopic calibration 
a posteriori by simulating quasar-galaxy lenses
and analysing them using the software actually used
for the survey.
This also has the advantage of taking into account all of the
potential biases in the survey without the need for explicit
modelling.

\subsection{Efficiency and completeness}
\label{section:e and c}

The sheer number of 
quasars that must be re-imaged per lens discovery in a conventional
lens survey
is prohibitive (\eg\ \sect{intro_grs}),
and it is one of the major reasons that so few lenses
are known. The use of morphological selection,
implicit in a \grs, can increase
the efficiency by removing most of the unlensed quasars from 
the survey, but there is also a significant reduction in 
the completeness of the lens sample.

The completeness is simply 
$C = 1 - p_{\rmn l}(< e_{\rmn min})|_{<m_{\rmn lim}}$, where $e_{\rmn min}$
is the ellipticity cut of the survey and $m_{\rmn lim}$ its
magnitude limit. 
As shown in \fig{p_e_lens}, this is quite variable,
with $C$ ranging between $0.05$ and $0.5$, depending mainly on the
form of the \psf.
However a low value of $C$ is not necessarily a problem, provided 
that its value is reasonably well known.
Note that $C$ is unambiguously defined only for the `elliptical'
lens sample; the lenses with $m_{\rmn g} \simeq \m_{\rmn q}$ 
are not drawn from a well-defined parent population.

\begin{figure}
\includegraphics{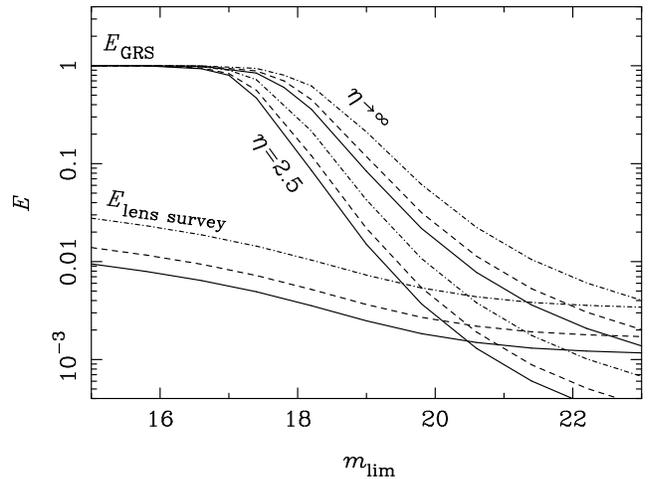}
\vspace{\singlefigureheight}
\caption{The detection efficiency of a
conventional lens survey, $E_{\rmn lens\,survey}$ (lower lines),
and that of a lens search based on a redshift
survey, $E_{\rmn \grs}$ (upper lines), 
as a function of the limiting magnitude 
of the survey, $m_{\rmn lim}$. Both are
defined in \sect{e and c}, and $E_{\rmn \grs}$ is shown 
for two values of $\eta$ used in the definition of the
Moffat (1969) \psf:
$\eta = 2.5$ and $\eta \rightarrow \infty$ (Gaussian seeing), as labelled.
In each case three cosmological models are used:
$\Omega_{\rm m_0} = 1.0$ and $\Omega_{\Lambda_0} = 0.0$ (solid lines);
$\Omega_{\rm m_0} = 0.3$ and $\Omega_{\Lambda_0} = 0.0$ (dashed lines);
and
$\Omega_{\rm m_0} = 0.3$ and $\Omega_{\Lambda_0} = 0.7$ (dot-dashed lines).}
\label{figure:eff_lens}
\end{figure}

More important is the efficiency, $E$, 
which is defined as the number of high-resolution re-observations
required per lens discovery.
For a redshift survey-based lens search it is 
[\cf\ \eq{eff}]
\begin{equation}
E_{\rmn \grs} = \frac{N_{\rmn l}(< m_{\rmn lim}) 
p_{\rmn l}(<e_{\rmn min})\,|_{<m_{\rmn lim}}}
{N_{\rmn q}(< m_{\rmn lim}) 
p_{\rmn q}(<e_{\rmn min})\,|_{<m_{\rmn lim}}},
\end{equation}
as only those objects with quasar-like spectra need to be re-imaged.
As shown in \fig{eff_lens}, this is generally
much higher than the efficiency of 
a conventional lens survey, 
\begin{equation}
E_{\rmn lens\,survey} = \frac{N_{\rmn l}(< m_{\rmn lim})}
{N_{\rmn q}(< m_{\rmn lim})},
\end{equation}
where it is optimisticly assumed that the lens sample is essentially complete.
For bright surveys $E_{\rmn \grs} \simeq 1$
as the star-galaxy separation techniques are so reliable,
although close quasar-galaxy associations have not been 
included. 
For fainter magnitude limits $E$ decreases for both search methods,
due to the lower lensing fraction, and for 
$m_{\rmn lim} \ga 20$ the 
efficiency of redshift surveys approaches that of lens surveys.

If only lenses with bright deflectors enter a galaxy survey,
then $E_{\rmn \grs}$ is reduced by about an order of magnitude,
but is still greater than $E_{\rmn lens\,survey}$ for 
$m_{\rmn lim} \la 20$.
Hence the efficiency of lens searches based on \grs\ spectra
is almost always greater than that of conventional lens surveys.

\subsection{Deflector redshift distribution}
\label{section:z_dist_grs}

\begin{figure}
\includegraphics{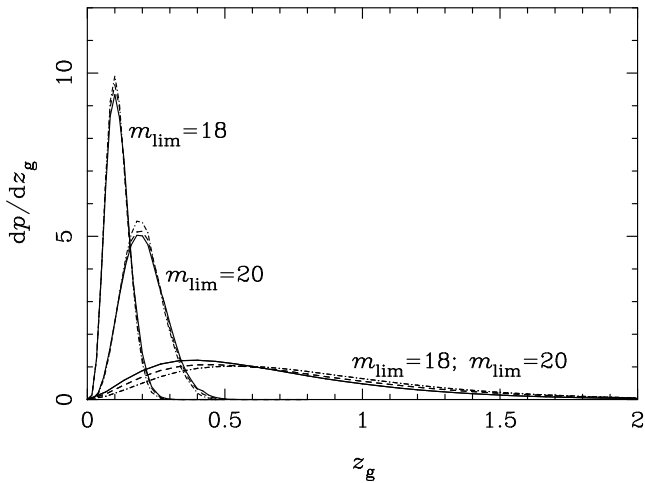}
\vspace{\singlefigureheight}
\caption{The normalised distribution of deflector redshifts, $z_{\rmn g}$,
of lensed quasars discovered in \grss. For the 
`elliptical' lenses (broader curves) the distribution is 
determined solely by the cosmological model, whereas the 
extent of the distribution for the lenses with bright deflectors 
increases with the depth of the survey. The two sets of 
curves for each type of 
lens show the distributions with $m_{\rmn lim} = 18$ and $m_{\rmn lim} = 20$.
For each survey  $\Delta m_{\rmn qg} = 2$, and
results are presented for three cosmological
models:
$\Omega_{\rm m_0} = 1.0$ and $\Omega_{\Lambda_0} = 0.0$ (solid lines);
$\Omega_{\rm m_0} = 0.3$ and $\Omega_{\Lambda_0} = 0.0$ (dashed lines);
and
$\Omega_{\rm m_0} = 0.3$ and $\Omega_{\Lambda_0} = 0.7$ (dot-dashed lines).}
\label{figure:dpdz_lens_grs}
\end{figure}

By evaluating the lensing probability (\sect{lensed quasars})
for a fixed deflector redshift,
the distribution of lens galaxy redshifts,
${\rmn d}p/{\rmn d}z_{\rmn g}$, can be found;
this is shown in \fig{dpdz_lens_grs}.
The `elliptical' lenses have the broad range of
deflector redshifts expected of a conventional lens survey
(\eg\ Turner \etal\ 1984),
and, in particular, the expected number of nearby
deflectors is very small.
The distribution is very different if
only lenses with visible lens galaxies are selected --
it roughly matches the overall redshift distribution
of a \grs\ with a limiting magnitude of $m_{\rmn lim} + \Delta m_{\rmn qg}$
(\cf\ \fig{dpdz_m_app}).
Hence the \TdF\ redshift survey should yield 
one or two lenses with $z_{\rmn g} \la 0.1$, 
and the \SDSS\ could contain about ten lenses with such 
nearby deflectors.

\section{Conclusions}
\label{section:conc}

Due to the imperfect discrimination between galaxies and 
other celestial objects, gravitationally-lensed quasars 
enter \grs\ catalogues, and should then be detectable 
spectroscopically. 
The gravitational lens Q~2237+0305 was discovered in 
this way (Huchra \etal\ 1985), and 
Kochanek (1992) predicted approximately one lens per $10^6$ redshifts
measured.
However the inclusion of the quasars' light in the calculation 
of the lens galaxies' magnitudes
increases the number of lenses by up to an order of magnitude,
as illustrated in \fig{delta_m_grs}. 
Another possibility is that many lensed quasars 
enter \grss\ as the quasar images combine to have a high
ellipticity, but are also unresolved
in the low-resolution (\eg\ plate) data
from which candidate galaxies are selected. 
If a significant fraction of such lenses are observed 
spectroscopically in \grss, more new lenses will be discovered 
in redshift surveys than are known to date.

The number of lenses  expected in various existing and planned
surveys is given in \tabl{n_lens_grs}.
It can be seen that $N_{\rmn l}$ varies greatly with both
the form of the \psf\ of the parent survey and
the cosmological model (in the case of the `elliptical' lenses),
which places fundamental limits on the accuracy of these predictions.
For instance,
the current generation of surveys (with up to
$\sim2 \times 10^4$ galaxies) should contain several 
lenses between them, but the above uncertainties, combined 
with shot noise, make this a rather weak prediction.
Looking ahead, the \TdF\ \grs\ (with $\sim 2.5 \times 10^5$ galaxies
to a limit of $m_{B_{\rmn J}} = 19.5$) is already well underway,
and should contain at least 10 new lenses, and up to 50 if 
observational conditions are favourable. 
Howver, it is particularly sensitive 
to surface brightness-related selection effects, 
and a more detailed simulation of lensing in the \TdF\ 
redshift survey is presented in Mortlock \& Webster (2000c).
Finally, the \SDSS\ can be expected to contain over 100 
spectroscopic lensed quasars, along with even more discovered
by more conventional methods.

Many, and possibly all, of the lenses discovered spectroscopically
in redshift surveys, will have deflector galaxies at low redshifts,
comparable to the depth of the survey proper. These have the 
potential to be the most important products of such a lens search,
as the proximity of the deflector, combined with the information
provided by the lensing event, can provide a number of unique insights, 
as exemplified by Q~2237+0305. 
The \TdF\ \grs\ should contain several low-redshift
lenses, and the \SDSS\ about 10. 
Of course not all of these will have Q~2237+0305's wonderful combination
of source and deflector properties, but the possibility of even 
one similar system is tantalising, to say the least.

\section*{Acknowledgments}

Matthew Colless, Paul Hewett and Steve Maddox 
went beyond the call of duty in answering an endless stream of 
e-mails about the the selection of objects into redshift surveys,
and Michael Drinkwater provided his data prior to publication.
DJM was supported by an Australian Postgraduate Award.

\bsp
\label{lastpage}
\end{document}